\documentclass[useAMS,usenatbib]{mn2e}
\usepackage{graphicx,amsmath,multirow,amssymb}
\usepackage{subfig}
\usepackage{natbib}
\newcommand{\comment}[1]{}

\def\simgt{\lower.5ex\hbox{$\; \buildrel > \over \F sim \;$}}
\def\simlt{\lower.5ex\hbox{$\; \buildrel < \over \sim \;$}}

\title[New post-AGB star models]{New Post-AGB star models as tools to understand AGB evolution and nucleosynthesis.}

\author[D. Kamath et al.]{
D. Kamath$^{1}$\thanks{E-mail: devika.kamath@mq.edu.au},
F. Dell'Agli$^{2}$, P. Ventura$^{2}$, H. Van Winckel$^{3}$, A. Karakas$^{4,5}$, S. Tosi$^{6}$
\\
$^{1}$Department of Physics and Astronomy, Macquarie University, Sydney, NSW 2109, Australia \\
$^{2}$INAF, Osservatorio Astronomico di Roma, Via Frascati 33, 00077, Monte Porzio Catone, Italy \\
$^{3}$Instituut voor Sterrenkunde, K.U.Leuven, Celestijnenlaan 200D bus 2401, B-3001, Leuven, Belgium \\
$^{4}$School of Physics \& Astronomy, Monash University, Clayton VIC 3800, Australia \\
$^{5}$RC Centre of Excellence for All Sky Astrophysics in 3 Dimensions (ASTRO 3D) \\
$^{6}$Dipartimento di Matematica e Fisica, Universit\'a degli Studi Roma Tre, via della Vasca Navale 84, 00100, Roma, Italy \\
}

\begin{document}

\date{Accepted, Received; in original form }

\pagerange{\pageref{firstpage}--\pageref{lastpage}} \pubyear{2012}

\maketitle

\label{firstpage}

\begin{abstract}
We study a sample of post-AGB stars in the Galaxy, with known surface chemical composition and
s-process enrichment. The recent determination of the luminosities of these sources, based on Gaia parallaxes, 
allows for the fist time a deep investigation of Galactic post-AGB stars, with the possibility of
characterising the individual objects in terms of mass, chemical composition and age of the
progenitors. To this aim we used available evolutionary sequences of AGB stars, extended to the post-AGB
phase, complemented by new models, specifically calculated for the present study. 
The combined knowledge of the surface carbon and of the luminosity proves the most valuable
indicator of the previous history of the star, particularly for what regards the efficiency
of the various mechanisms able to alter the surface chemical composition and the growth of the core mass.
This kind of analysis allows dissecting different classes of stars, such as low-mass objects, that evolved
as M-type stars for the whole AGB lifetime, carbon stars, massive AGB stars that experienced hot
bottom burning. The potentialities of this approach to shed new light on still debated issues related
to the AGB evolution are also commented.
\end{abstract}




\section{Introduction}
The stars evolving across the asymptotic giant branch (AGB) have attracted the interest of
the astrophysical community in the last decades, owing to the relevant role played in different
astrophysical contexts, such as the interpretation of the chemical patterns traced by stars
in the Milky Way \citep{romano10} and other galaxies \citep{vincenzo16, romano20}, the formation of multiple
populations in globular clusters \citep{dercole08, dercole10}, the derivation of the mass of
high-redshift galaxies \citep{claudia06}. Furthermore, AGB stars are generally regarded as important
dust manufactures, providing a relevant contribution to the overall dust budget in the Universe,
in the Local Universe and at high redshift \citep{valiante09}.

The modelling of the AGB phase has been significantly improved in recent times, with a more
physically sound description of some physical mechanisms, such as the computation of low-temperature 
opacities in environments enriched in carbon \citep{marigo02} and a self-consistent coupling of
nuclear burning and mixing of chemicals in regions of the star unstable to convection
\citep{herwig00, herwig05}. More recent efforts were devoted to include the description of
dust formation in the AGB winds \citep{ventura12, nanni13, nanni14, ventura14}, which allowed
the study of evolved stellar populations in the Magellanic Clouds \citep{flavia14, flavia15a, flavia15b, 
nanni16, nanni18, nanni19b} and other galaxies in the Local Group \citep{flavia16, flavia18b, flavia19a}.

Despite the recent improvements, the robustness of the results provided by the modelling of the
AGB phase is hampered by the poor knowledge, from first principles, of some physical mechanisms
that deeply affect the internal structure and the evolution of AGB stars, primarily convection
and mass loss \citep{karakas14}. While non-local models of the convective instability 
and self-consistent descriptions of mass-loss are being developed, the comparison with the observational 
evidence is presently the only possibility to provide valuable indications of the efficiency of some physical 
mechanisms, particularly those affecting directly the surface chemical composition of the stars. 
Indeed the knowledge of the surface chemistry of evolved stars proves extremely useful to 
their characterisation, as it is extremely sensitive to the balance between 
various mechanisms, such as hot bottom burning (HBB, Bl\"ocker \& Sch\"onberner 1991) and third dredge-up
(TDU, Iben 1974), which would produce different chemical patterns: the former favours an enrichment
in the surface $^{12}$C, whereas HBB changes the mass distributions among the various elements
according to the equilibria of the proton-capture nucleosynthesis activated at the bottom of the
convective envelope.

A valuable approach to infer information on the evolution of AGB stars is offered by 
the study of post-AGB stars. The chemical composition of these objects reflects the final surface 
chemistry, at the end of the AGB phase, and is determined by the combination of TDU and HBB. 
On the observational side, determination of the surface chemical composition of post-AGB stars
is easier than for AGBs, because the optical/near-IR spectra of such cool giants are contaminated by millions 
of molecular lines, and deriving the abundances of individual species requires the use of spectral synthesis
techniques (e.g. Garc{\'{\i}}a Hern{\'a}ndez et al. 2006, 2009), even when taking into account circumstellar 
effects. In addition, the most extreme AGBs are heavily obscured and escape detection (and abundance studies) 
in the optical range (e.g. Garc{\'{\i}}a Hern{\'a}ndez et al. 2007) and/or they may display extremely complex 
near-IR spectra (e.g. McSaveney et al. 2007). 

An additional point in favour of studying post-AGB objects is that the excursion to the blue, towards the
planetary nebulae phase, occurs at a substantially constant luminosity, which depends on the core mass
at the end of the AGB. The latter quantity is tightly connected with mass and metallicity of the progenitor, 
which allows the comparison between the observed surface abundances of the individual chemical species and the
expectations from stellar evolution modelling, based on the description of the above mentioned mechanisms
able to change the surface chemistry, mainly TDU and HBB. Therefore, the determination of the luminosity of 
post-AGB stars, combined with the knowledge of their surface chemical composition, allows not only to study
the details of the post-AGB phase, but also to draw important information on the previous AGB evolution. 

This approach has been so far only partially applied to Galactic post-AGB stars, owing to the poor
knowledge of their distances, which has prevented an accurate estimate of their luminosities. Results 
from Gaia are partly removing this limitation, by providing distances and luminosities from a growing
sample of post-AGB stars, with known chemical composition. The combined information
of luminosity and chemical composition of a sample of Galactic post-AGB sources have been recently 
published by Kamath et al (2021, hereafter paper I), who presented the mass fractions of CNO elements and s-process
enrichment for 31 post-AGB sources, with available metallicity and luminosity. 

The goal of the present work is to study the sample of stars presented in paper I, by comparing results from 
observations with predictions from AGB evolution modelling. We use AGB models by our group, already available 
in the literature, complemented by new evolutionary sequences, extended to the post-AGB phase, when required. 
This study is the first step towards the development of a methodology aimed at the interpretation and the 
characterization of post-AGB stars observed in the Galaxy, to reconstruct their evolutionary hystory, opening 
the way to the use of these stars to draw information on the chemical evolution, the metal enrichment and the 
star formation history of their host system, even external to the Milky Way.

\section{The chemical composition of AGB stars, on their path to the post-AGB phase}
\label{chem}
The evolution of the stars through the AGB phase is extensively described in dedicated reviews on the argument, such
as those by \citet{herwig05} and more recently by \citet{karakas14}. A peculiar feature of this evolutionary
phase is the occurrence of a series of thermal pulses (TP), taking place periodically, when helium burning is activated in a stellar region above the degenerate core under conditions of thermal instability \citep{sh65}. After each TP the bottom of the surface convective region penetrates inwards, until crossing, under specific conditions, the H$/$He discontinuity, thus entering regions of the star previously exposed to $3\alpha$ nucleosynthesis, greatly enriched in $^{12}$C. This mixing episode is commonly referred to as third dredge-up \citep{iben74, mowlavi99}. The occurrence of TDU starts after a few TPs since the beginning of the AGB phase, and favours a gradual enrichment in the surface $^{12}$C content and, to a lower extent, of $^{16}$O. This is accompanied by the surface s-process enrichment \citep{busso91}.

All the stars experiencing TDU during the AGB evolution are expected to show up large $^{12}$C enhancement and s-process enrichment during the post-AGB phase. No detection of carbon enrichment in post-AGB stars can be due to the following reasons: a) the mass of the envelope at the beginning of the AGB was so small that the whole external mantle was lost after only a few TPs, with no (or scarce) effects of TDU; b) the star experienced an early loss of the envelope, owing to e.g. a tight gravitational encounter or to mass transfer to a companion star; c) the star experienced HBB, with the depletion of the surface $^{12}$C via proton capture reactions taking place in the innermost regions of the envelope, once the temperatures in those zones reach (or exceeed) $\sim 30$ MK.

Case (a) is associated to the evolution of low-mass stars, with total mass at the beginning of the AGB of the order of $\sim 1~\rm{M}_{\odot}$, or below. These stars are made up of a degenerate core, of mass $\sim 0.5~\rm{M}_{\odot}$, composed of carbon and oxygen, and a $\sim 0.5~\rm{M}_{\odot}$ envelope, which is lost with a typical time-scale of $\sim 1$ Myr, after the star experienced $4-5$ TPs, with no TDU. This class of objects is expected to enter the post-AGB phase with luminosities below $\sim 5000~\rm{L}_{\odot}$ \citep{marcelo16}, and to exhibit poor (if any) $^{12}$C enrichment and $^{14}$N enhancement, as a consequence of the first dredge-up (hereafter FDU, see e.g. Sweigart et al. 1989) and possibly of deep mixing during the RGB ascending \citep{lagarde12, lagarde19}. The mass threshold below which the stars follow this kind of evolution is sensitive to the metallicity of the star and is still debated, as the modelling of TDU is extremely sensitive to the treatments of the convective borders, the related mixing at the base of the convective envelope, and the description of mass loss \citep{karakas14}. 

Case (b) corresponds to stars which undergo a fast loss of the external envelope, so that the AGB evolution is accelerated with the rapid contraction of the structure, which starts the post-AGB phase. The number of TPs experienced, and also of the TDU events, is lower the earlier is the loss of the external mantle. The growth of the core mass and luminosity is also truncated.

\begin{figure*}
\begin{minipage}{0.48\textwidth}
\resizebox{1.\hsize}{!}{\includegraphics{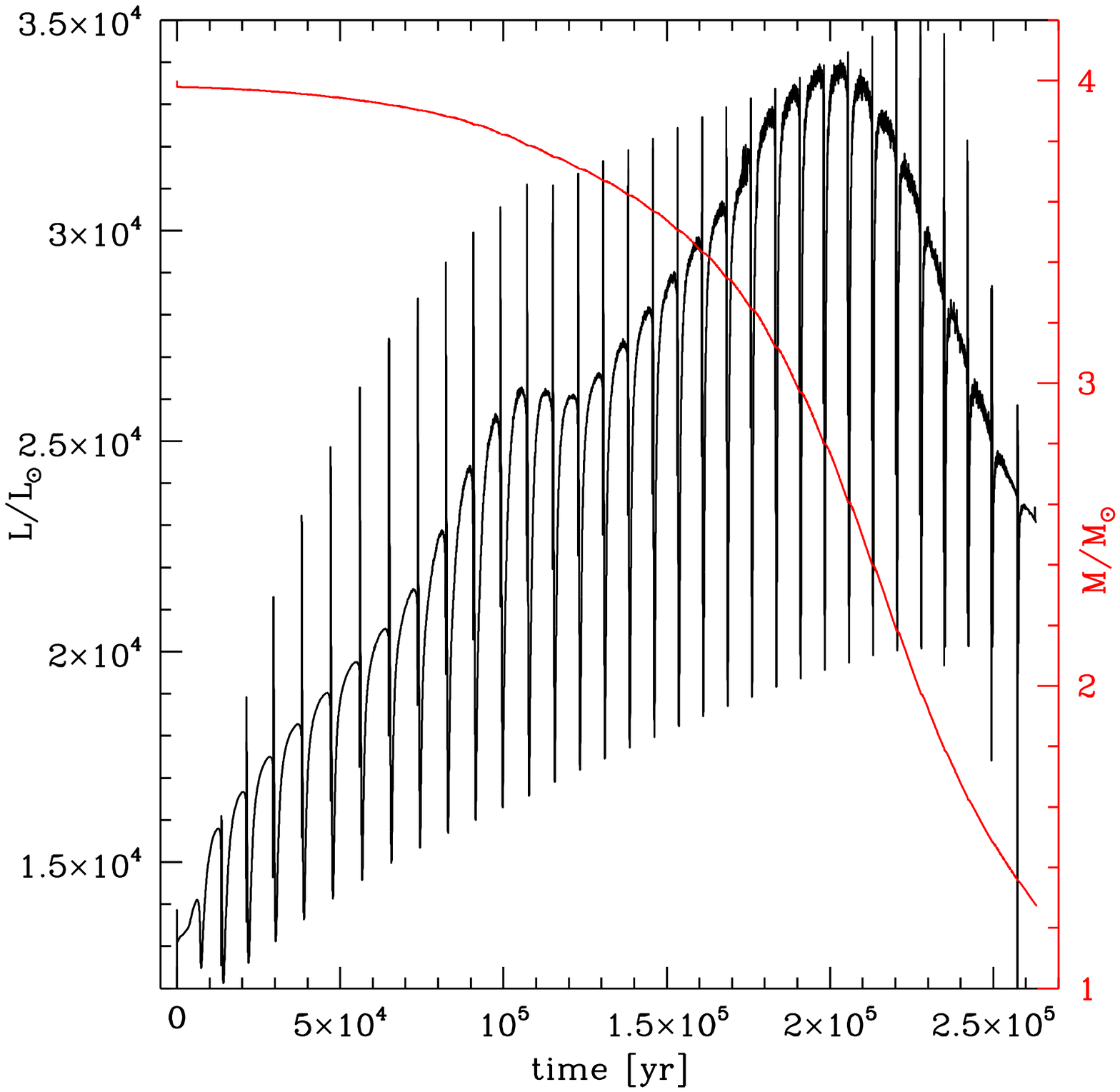}}
\end{minipage}
\begin{minipage}{0.48\textwidth}
\resizebox{1.\hsize}{!}{\includegraphics{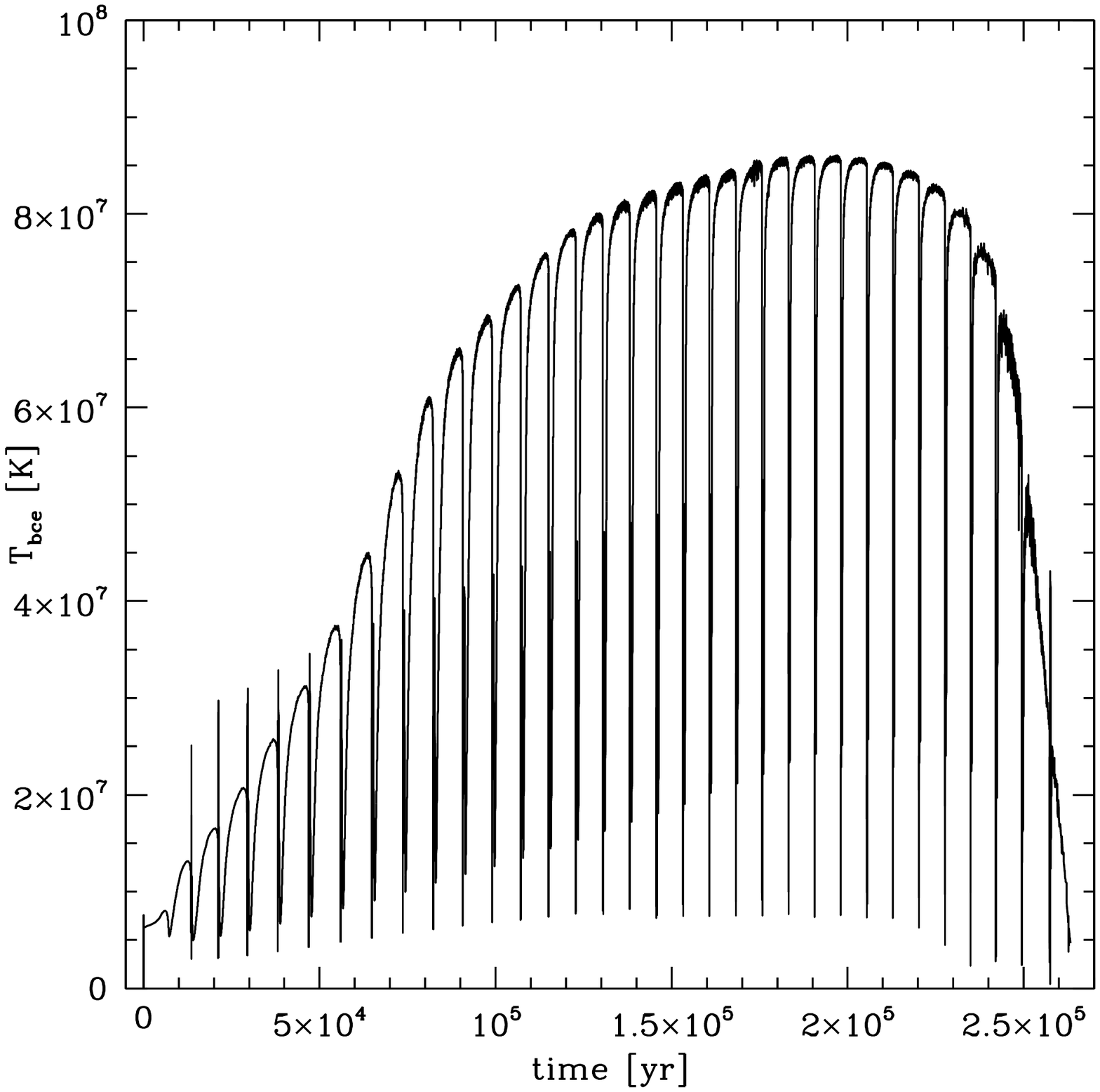}}
\end{minipage}
\vskip-80pt
\begin{minipage}{0.48\textwidth}
\resizebox{1.\hsize}{!}{\includegraphics{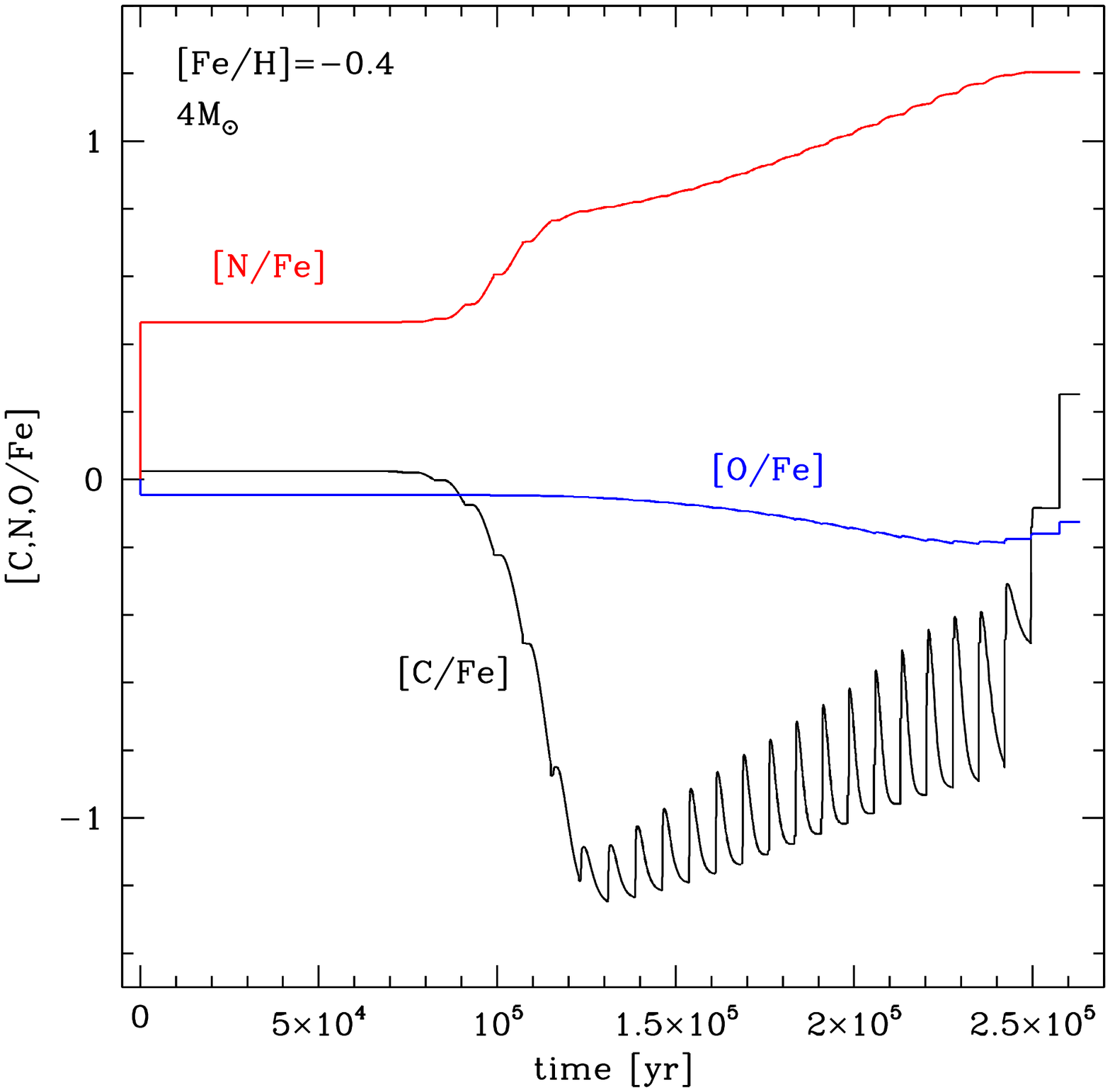}}
\end{minipage}
\begin{minipage}{0.48\textwidth}
\resizebox{1.\hsize}{!}{\includegraphics{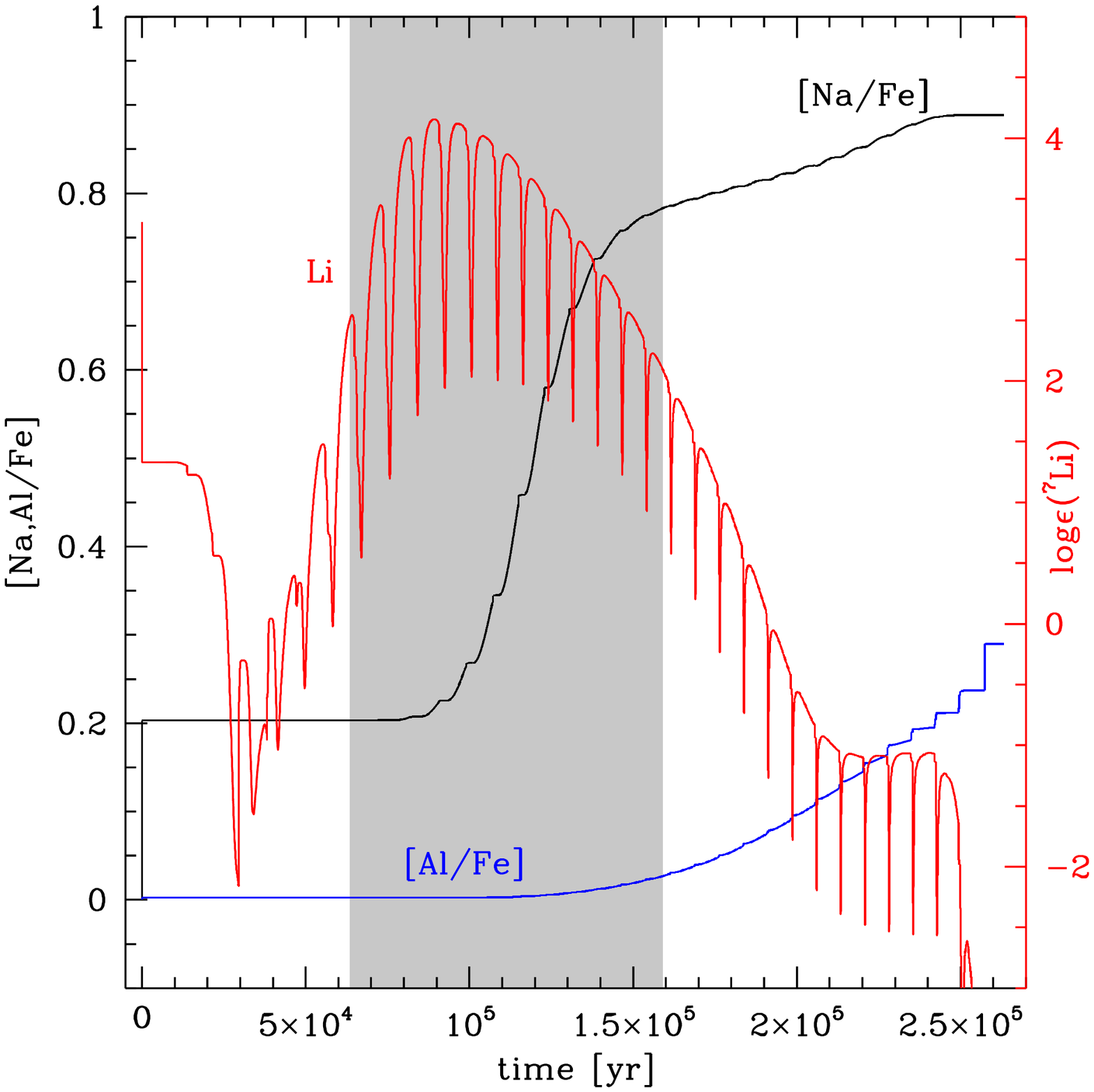}}
\end{minipage}
\vskip-60pt
\caption{Top, Left: AGB variation of the luminosity (black line, scale on the left y-axis) and current mass (red, scale on the right) of a $4~\rm{M}_{\odot}$, $Z=8\times 10^{-3}$ 
model star, experiencing HBB. Top, Right: Run of the temperature at the base of the envelope. The fast excursions
of the luminosity and temperature at the base of the envelope correspond to the ignition of TPs.The bottom panels report the variation of the surface CNO mass fractions (bottom, left) and of magnesium, aluminium and lithium
(bottom, right). The scale for lithium, in red, is on the right vertical axis and is given in the
$\log \epsilon (^7{\rm Li})=12+\log({\rm n}(^7{\rm Li})/{\rm n}({\rm H})))$ unit. Grey shading indicates the super-rich lithium phase, when $\log \epsilon (^7{\rm Li})>2$.}
\label{fhbb}
\end{figure*}

The stars corresponding to case (c) above are those which evolve on core masses above $0.8~\rm{M}_{\odot}$, which reflects into initial masses $\rm{M}\geq 3~\rm{M}_{\odot}$ \citep{ventura13}. The activation of HBB is connected with the steep gradients of the thermodynamic variables (primarily temperature and density) in the regions between the outer border of the degenerate core and the internal regions of the envelope, which determines temperatures at the base of the external mantle sufficiently hot to favour advanced proton-capture nucleosynthesis, whose products are transported to the surface by fast convective currents. The stars undergoing HBB evolve at large luminosities during the AGB lifetime, spanning the $2\times 10^4~\rm{L}_{\odot}-10^5~\rm{L}_{\odot}$ range, which typically reach a peak value, before the gradual loss of the envelope makes the luminosity to decrease, and to enter the post-AGB phase with values in the $1.5\times 10^4~\rm{L}_{\odot}-3\times 10^4~\rm{L}_{\odot}$ range. 

A typical example of such a behaviour can be seen in the top, left panel of Fig.~\ref{fhbb}, that shows the evolution of the luminosity of a $4~\rm{M}_{\odot}$ model star: we notice the fast increase in the luminosity, which grows up to $\sim 35000~\rm{L}_{\odot}$, then declines during the final AGB phases, after the total mass has decreased by $\sim 50\%$. A similar trend is followed by the temperature at the bottom of the envelope, $\rm{T}_{\rm bce}$, which in this specific case grows up to $\sim 80$ MK, in conjunction with the luminosity peak, then declines until HBB is turned off.

The bottom panels of Fig.~\ref{fhbb} show the evolution of the surface chemical composition. On the left we report the mass fractions of the most relevant chemicals entering the CNO cycling. The surface chemical composition of post-AGB stars previously exposed to HBB is significantly enriched in nitrogen, as CNO nuclearly processed matter is transported to the external regions of the star \citep{karakas14, karakas16, ventura02, ventura13}: we can distinguish in Fig.~\ref{fhbb} the effects of FDU, in the initial rise in the surface N, up to $[$N$/$Fe$] \sim 0.5$, and of HBB, starting after $\sim 8\times 10^4$ yr, when the surface carbon is depleted. In the case reported in Fig.~\ref{fhbb} the combined effects of FDU and HBB lead to a surface nitrogen enrichment of a factor $\sim 20$ with respect to the initial mass fraction. 

The HBB temperatures reached by the model star shown in
Fig.~\ref{fhbb} are sufficiently large to start the Ne-Na nucleosynthesis, with the production of sodium. These conditions are not sufficient to the destruction of the surface $^{24}$Mg, but are hot enough for the proton capture reactions by the two less abundant isotopes of magnesium, $^{25}$Mg and $^{26}$Mg: the overall magnesium will remain substantially unchanged, whereas some synthesis of $^{27}$Al occurs. This is shown in the bottom, right panel of Fig.~\ref{fhbb}, where we see that both sodium and aluminium increase to $[$Na$/$Fe$]\sim 1$ and $[$Al$/$Fe$]\sim 0.3$, respectively. 

We also see in the same panel the effects of the activation of the \citet{cameron} mechanism, which favours the synthesis of large quantities of lithium, as a consequence of a series of reactions started by
$\alpha$-captures by $^3$He nuclei.; the lithium-rich phase, outlined with grey shading in Fig.~\ref{fhbb}, lasts until there is some $^3$He available in the envelope, then it is destroyed, until the envelope is substantially lithium-free. Whether the stars experiencing HBB enter the post-AGB phase with some lithium in the surface regions depends on the strength of the HBB experienced, thus the largest temperatures reached at the bottom of the convective envelope. For $\rm{T}_{\rm bce}$ in excess of $\sim 60$ MK the consumption of the surface $^3$He is so fast that the lithium-rich phase is extremely short, thus preventing the survival of lithium until the beginning of the post-AGB evolution; this is the case shown in Fig.~\ref{fhbb}. Conversely, when HBB is weaker, lithium survives until the end of the AGB evolution.

The surface fraction of helium is also expected to increase in $\rm{M} \geq 4~\rm{M}_{\odot}$ stars with respect to the initial quantity, owing to the effects of the second dredge-up \citep{boothroyd99}
In metal-poor environments it is expected a reduction of the surface oxygen, whose extent is extremely sensitive to the metallicity \citep{flavia18a}; a small oxygen depletion is indeed visible in Fig.~\ref{fhbb}. 

The predictions regarding the carbon abundance for the stars experiencing HBB are more tricky, as the surface
$^{12}$C is sensitive to the balance between the effects of HBB, which destroys the surface $^{12}$C via proton fusion, and TDU, which brings freshly synthesised carbon to the surface regions. According to the modelling used in the present investigation TDU and HBB approximately balance each other for stars in the $3-4~\rm{M}_{\odot}$ mass range, whereas in higher mass stars strong HBB quenches TDU, thus preventing both carbon enrichment and s-process enhancement \citep{boothroyd93}. The effects of TDU and HBB are clear in Fig.~\ref{fhbb}, particularly those associated to late TDU events, which rise the surface $^{12}$C.
However the argument is still highly debated because both the strength of HBB and the inwards penetration of the convective envelope after each TP are extremely sensitive to the description of the convective instability \citep{karakas14}.

Regarding the surface abundances of the individual chemical species, and the predictions regarding the
mass fractions during the post-AGB phase, we focus on the CNO elements and neglect helium in the following discussion, because the helium abundances are not available for the sample stars discussed in paper I. 

\begin{figure*}
\begin{minipage}{0.48\textwidth}
\resizebox{1.\hsize}{!}{\includegraphics{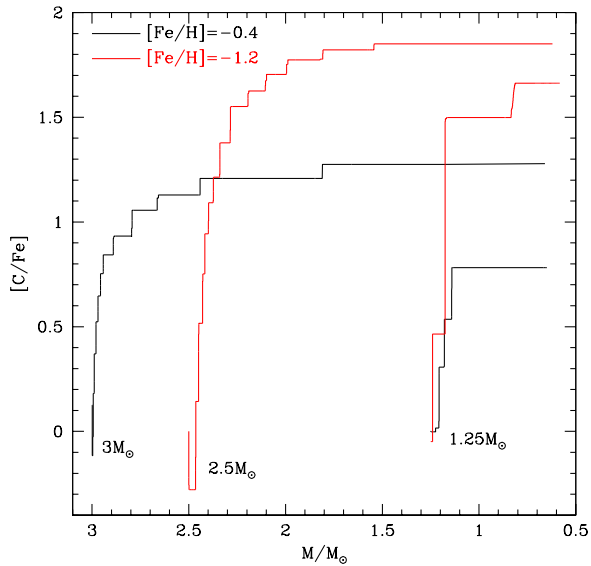}}
\end{minipage}
\begin{minipage}{0.48\textwidth}
\resizebox{1.\hsize}{!}{\includegraphics{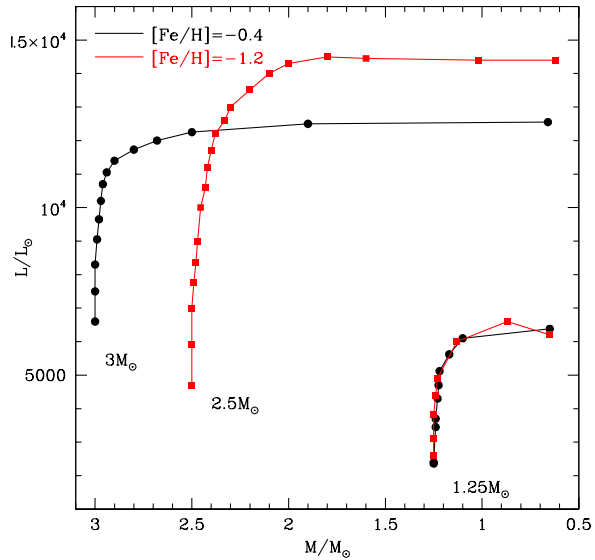}}
\end{minipage}
\vskip-60pt
\caption{AGB variation of the surface carbon (left panel) and luminosity (right) of model stars of different mass and metallicity, undergoing TDU. On the abscissa we report the current mass of the star. The points on the tracks on the right panel correspond to the inter-pulse phases.}
\label{fdup}
\end{figure*}

For the stars not included in cases (a), (b) and (c) above, descending from progenitors with mass in the $1-3~\rm{M}_{\odot}$ range, we expect to observe significant carbon enrichment, owing to the effects of repeated TDU events. A few example of the evolution of these objects can be seen in Fig.~\ref{fdup}, that reports the AGB variation
of the surface carbon and the luminosity for stars of different mass and metallicity. The masses were selected so that both the stars that experience only a few TDU events and those that attain the largest carbon enrichment are
represented in the figure: for a given metallicity, the final $[$C$/$Fe$]$ is higher the higher the initial
mass of the star.

An important result, clearly visible in Fig.~\ref{fdup}, is that the final $[$C$/$Fe$]$ is higher the lower is the metallicity. This is partly due to the higher efficiency of TDU experienced by lower metallicity AGBs, as suggested by
observations of external galaxies such as the LMC, SMC, and dwarf spheroidal galaxies, that show higher numbers of C-stars than in the Milky Way \citep{albert06, sloan12}, which is evidence that it is easier to obtain TDU at lower metallicities, for a given mass. However, the most relevant effect to consider on this regard is that, for a given amount of carbon transported to the surface by TDU, the percentage increase in the surface carbon with respect to the initial quantity is higher the lower is the initial mass fraction of $^{12}$C, which scales with the metallicity. 
On regard of the results of Fig.~\ref{fdup}, the model stars with initial mass $2.5~\rm{M}_{\odot}$ and
$3~\rm{M}_{\odot}$, with final $[$C$/$Fe$]$ $\sim 1.85$ and $\sim 1.3$, respectively, enter the post-AGB phase
with similar surface carbon mass fractions, slightly below 1.5.

This will be important in the interpretation of the carbon data, which will be addressed in the following sections. 
These stars are also expected to exhibit nitrogen enhancement, as a result of RGB mixing. Some oxygen enrichment
is possible if the inwards penetration of the surface convection during the TDU events is sufficiently
deep to reach layers enriched in $^{16}$O. We will return to this point in the next section.

The stars discussed in point (a) above are expected to show-up chemical compositions reflecting the effects of the FDU, with a possible further contribution from 1 or 2 TDU episodes. The occurrence of the FDU favours the surface nitrogen enrichment and the depletion of the surface carbon. If no TDU episode takes place, the surface carbon is characterized by negative values of $[$C$/$Fe$]$. However, a single TDU event is sufficient to restore solar-scaled or slightly higher values. Oxygen is untouched by FDU, thus it is expected that these stars reach the post-AGB with the same oxygen as in the gas from which they formed.

Finally, for what concerns the stars treated in point (c), which experienced HBB, the most clear signature of the nuclear activity at the base of the envelope, as clear in Fig.~\ref{fhbb}, is the surface nitrogen enhancement. This expectation is rather robust, as it is substantially independent of the strength of HBB, since CN cycling is the nuclear activity most easily activated
under HBB conditions. This is different from other nuclear channels, such as Ne-Na cycling or the Mg-Al-Si chain,
whose ignition critically depend on the temperature at the bottom of the convective envelope (see e.g. Ventura et al. 2018 for a more exhaustive discussion on the sensitivity of the degree of the HBB nucleosynthesis to convection modelling in solar metallicity AGBs). In case that the modification of the surface chemistry is determined by HBB only, with no effects related to TDU, the expected increase in $[$N$/$Fe$]$ is not sensitive to the metallicity, as  the nitrogen produced is of secondary origin, being synthesised by the CNO elements present in the mixture from which the stars formed. In metal-poor stars significant depletion of the surface oxygen takes place \citep{flavia18a}, but the metallicities of the sample stars discussed here are not sufficiently low to see these effects.

\section{AGB and post-AGB modelling}
\label{models}
The present work is based on AGB models of stars of different mass, spanning the metallicities $10^{-3} < Z < 0.014$, approximately corresponding to the $-1.5 < [\rm{Fe}/\rm{H}] < 0$ range, which allows the study of all
the sources in the paper I sample. These AGB models were presented in previous works, namely
\citet{ventura14} for $Z=0.001$, $Z=0.002$, $Z=0.004$, \citet{marini21} for $Z=0.008$ and
\citet{ventura18} for $Z=0.014$. The interested reader is addressed to these papers for the details
on the physical input used to calculate the evolutionary sequences and the main physical and chemical properties of the model stars. We give here only the following assumptions adopted, most relevant for the present work: a) the temperature gradient within regions of the star unstable to convection are found vis the
full spectrum of turbulence (hereafter FST) model \citep{cm91}; b) extra-mixing is assumed from convective borders during the AGB phase, by assuming an exponential decay of convective velocities, with an e-folding distance of $0.02~\rm{H}_{\rm p}$, the latter quantity being evaluated at the formal border, fixed via the Schwarzschild criterion; c) the computation of the surface molecular opacities takes into account changes in the mixture connected to carbon enrichment, according to the schematisation by \citet{marigo09}.

\begin{table}
\caption{The post-AGB luminosity (col. 2), core mass (3) and surface CNO mass fractions (col. 4, 5, 6) of 
model stars of different mass (col. 1) and metallicity. The initial masses refer to the beginning of the
AGB phase.}                                       
\begin{tabular}{c c c c c c c}    
\hline
$\rm{M}/\rm{M}_{\odot}$  & L/L$_{\odot}$  & $\rm{M}_{\rm core}/\rm{M}_{\odot}$ & $[$C$/$Fe$]$ &  $[$N$/$Fe$]$ & $[$O$/$Fe$]$  \\
\hline
 & & Z=$10^{-3}$ & & \\
\hline
0.70 & 3850 & 0.55  & -0.04 & 0.55 & 0.40 \\
0.75 & 4100 & 0.565 & -0.04 & 0.56 & 0.40 \\
0.80 & 4300 & 0.561 &  0.80 & 0.56 & 0.40 \\
0.85 & 4500 & 0.565 &  0.82 & 0.57 & 0.40 \\
0.95 & 5050 & 0.567 &  1.50 & 0.57 & 0.51 \\
1.25 & 7000 & 0.58  &  2.45 & 0.58 & 0.60 \\
1.50 & 7600 & 0.593 &  2.12 & 0.59 & 0.80 \\
\hline
 & & Z=$2\times 10^{-3}$ & & \\
\hline
0.75 & 4000 & 0.560 & -0.05 & 0.57 & 0.40 \\ 
0.77 & 4700 & 0.564 &  0.00 & 0.57 & 0.40 \\
0.80 & 4800 & 0.562 &  1.01 & 0.57 & 0.40 \\
0.90 & 5300 & 0.565 &  1.06 & 0.57 & 0.40 \\
1.00 & 5400 & 0.57  &  1.30 & 0.58 & 0.45 \\
1.10 & 5900 & 0.572 &  1.51 & 0.58 & 0.49 \\
1.25 & 6500 & 0.584 &  1.65 & 0.58 & 0.50 \\
\hline
 & & Z=$4\times 10^{-3}$ & & \\
\hline
0.80 & 4500 & 0.564 & -0.02 & 0.10 & 0.20 \\
0.85 & 5250 & 0.571 &  0.00 & 0.12 & 0.20 \\
0.90 & 5100 & 0.564 &  0.83 & 0.15 & 0.20 \\
1.00 & 5400 & 0.566 &  0.92 & 0.21 & 0.20 \\
1.25 & 5900 & 0.573 &  0.98 & 0.30 & 0.26 \\
1.50 & 6700 & 0.582 &  1.09 & 0.36 & 0.29 \\
1.75 & 7800 & 0.588 &  1.25 & 0.41 & 0.29 \\
2.00 & 7850 & 0.60  &  1.25 & 0.48 & 0.38 \\
2.50 &10100 & 0.63  &  1.51 & 0.53 & 0.50 \\
3.00 &16050 & 0.75  &  1.32 & 0.47 & 0.55 \\
3.50 &19000 & 0.84  &  0.63 & 1.52 & 0.30 \\
\hline
 & & Z=$8\times 10^{-3}$ & & \\
\hline
0.75 & 4000 & 0.55  &  0.0  & 0.30 & 0.20 \\
0.80 & 4200 & 0.557 &  0.0  & 0.30 & 0.20 \\
0.85 & 5100 & 0.564 &  0.0  & 0.30 & 0.20 \\
0.90 & 5550 & 0.574 &  0.0  & 0.30 & 0.20 \\ 
0.95 & 5500 & 0.571 &  0.58 & 0.30 & 0.20 \\
1.00 & 6200 & 0.584 &  0.44 & 0.32 & 0.22 \\
1.25 & 6500 & 0.59  &  0.60 & 0.30 & 0.23 \\
1.50 & 7450 & 0.602 &  0.59 & 0.36 & 0.24 \\
1.75 & 8300 & 0.632 &  0.69 & 0.39 & 0.20 \\   
2.0  & 8200 & 0.617 &  0.80 & 0.44 & 0.20 \\
2.5  & 8600 & 0.606 &  1.12 & 0.55 & 0.26 \\
3.0  & 9300 & 0.66  &  1.14 & 0.57 & 0.50 \\ 
3.5  &21500 & 0.79  &  0.35 & 1.55 & 0.36 \\
4.0  &23000 & 0.87  &  0.22 &  & \\
\hline
\label{tabpost}
\end{tabular}
\end{table}

All the AGB computations mentioned above were extended until the very end of the AGB phase, when the mass of
the envelope was reduced to below a few tenths (in some cases a few hundredths) of solar masses; the only exception are the $Z=0.008$ models published in \citet{marini21}, which are extended to the post-AGB phase. The present
work is aimed at the interpretation of a sample of post-AGB stars, which requires the determination
of the luminosity and the surface chemistry expected during the post-AGB evolution. For the chemical
composition we might rely with sufficient accuracy to the surface mass fractions of the individual
species found at the end of the AGB lifetime, as no further modification of the surface chemistry is expected
since the phase until which the computations are extended; the only exception to this is the occurrence of a late TP, which induces a further alteration of the chemical composition of the external regions of the star
\citep{herwig99}. For what concerns the luminosity, the question is more tricky, as the detailed modelling of the transition phase from the AGB to the post-AGB is required to
predict the post-AGB luminosity with sufficient accuracy. For these reasons, specifically for the present investigation, the afore-mentioned evolutionary
sequences were extended to the post-AGB phase. Note that this was done only for model stars with specific
initial mass, whose luminosities are consistent with the values derived in paper I; in the very low-mass domain
this required the computation of new evolutionary sequences, not included in the papers cited above.

A summary of the evolutionary results from the extension of the AGB computations is reported in 
tab.~\ref{tabpost}, giving the post-AGB luminosities, core masses and the surface abundances of
the CNO species of model stars of various mass and metallicity. Note that for $\rm{M} \leq 2~\rm{M}_{\odot}$ stars, undergoing the helium flash, the masses reported in col.~1 refer to the values at the beginning of the
core helium burning phase, thus neglecting the mass loss taking place during the RGB ascending: therefore, 
the masses reported in col.~1 are a lower limit of the mass of the progenitors, particularly for the
stars in the $\rm{M} < 1.5~\rm{M}_{\odot}$ mass domain, where we expect a RGB mass loss of the order
of $0.1-0.2~\rm{M}_{\odot}$.

Fig.~\ref{fclum} shows the surface carbon and luminosity with which the model stars evolve through the
post-AGB phase: the results corresponding to different masses are indicated with filled circles, those
of same metallicity being connected with solid lines; the colour-coding corresponds to the metallicity, as indicated in the right panel of Fig.~\ref{fclum}.

On this plane the evolutionary sequences trace an anticlockwise trend with mass. The tracks first move
to the right, as the higher the initial mass of the star, the higher the number of TP and TDU events
experienced, the larger the surface carbon enrichment. There is little increase in the luminosity in the
initial part of the tracks, because $\rm{M} \leq 2~\rm{M}_{\odot}$ stars develop
electron degeneracy in the central regions during the RGB evolution and start the AGB phase with similar core masses. The left turn of the tracks for masses $\sim 2.5-3~{\rm M}_{\odot}$ is connected with the activation of HBB, which destroys the surface carbon previously accumulated via TDU. In this
case the luminosity is more sensitive to the mass of the star, because the core masses are different
and the strength of the HBB experienced, thus the consequent increase in the luminosity, change significantly
with the mass of the star.

The nitrogen abundances reported in tab.~\ref{tabpost} are $[$N$/$Fe$]>0$ in all cases, with most of the model
stars reaching the post-AGB phase with $[$N$/$Fe$] \sim 0.3-0.5$. This is mostly related to the effects of
FDU, which rises the surface nitrogen of the star. Note that
the N values reported in tab.~\ref{tabpost} must be considered as lower limits, as no RGB deep mixing related
to e.g. thermohaline mixing \citep{eggleton06} or rotation \citep{charbonnel95} was considered in the models used in the present investigations. $\rm{M} \geq 3.5~\rm{M}_{\odot}$ model stars are characterised by large N abundances $[$N$/$Fe$]>1$, a signature of the HBB experienced during the AGB.

From the oxygen abundances reported in tab.~\ref{tabpost} we see that no change in the surface $^{16}$O
is expected in $\rm{M} < 1~\rm{M}_{\odot}$ stars, as FDU leaves the surface oxygen
unchanged, and the TDU episodes are too few (if any) and in any case not sufficiently deep to trigger
a modification in the surface oxygen. Higher mass stars experience a higher number of TDU events,
sufficiently deep to reach regions of the stars enriched in $^{16}$O, which leads to an increase in the
post-AGB $[$O$/$Fe$]>0$. The largest oxygen enhancement, $\delta [$O$/$Fe$] \sim 0.3$, are found in $\sim 3~\rm{M}_{\odot}$ stars, which experience the largest number of TDU events.

\section{The characterisation of post-AGB stars}
\label{interp}
We characterise the individual post-AGB sources in the sample presented in paper I by comparing the observed chemical composition and the derived luminosities with the results from stellar evolution modelling presented in the previous section.

\begin{figure*}
\begin{minipage}{0.48\textwidth}
\resizebox{1.\hsize}{!}{\includegraphics{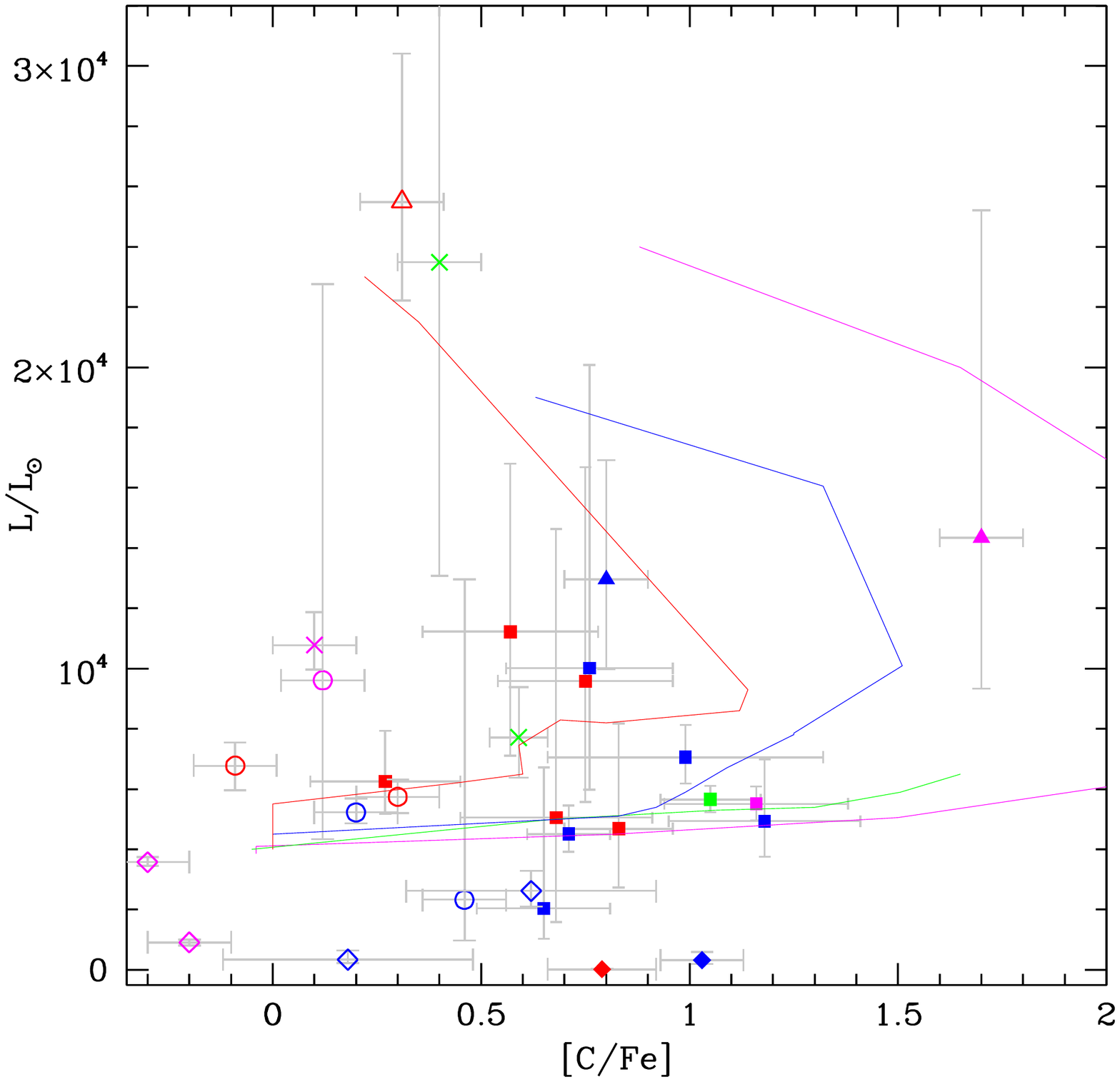}}
\end{minipage}
\begin{minipage}{0.48\textwidth}
\resizebox{1.\hsize}{!}{\includegraphics{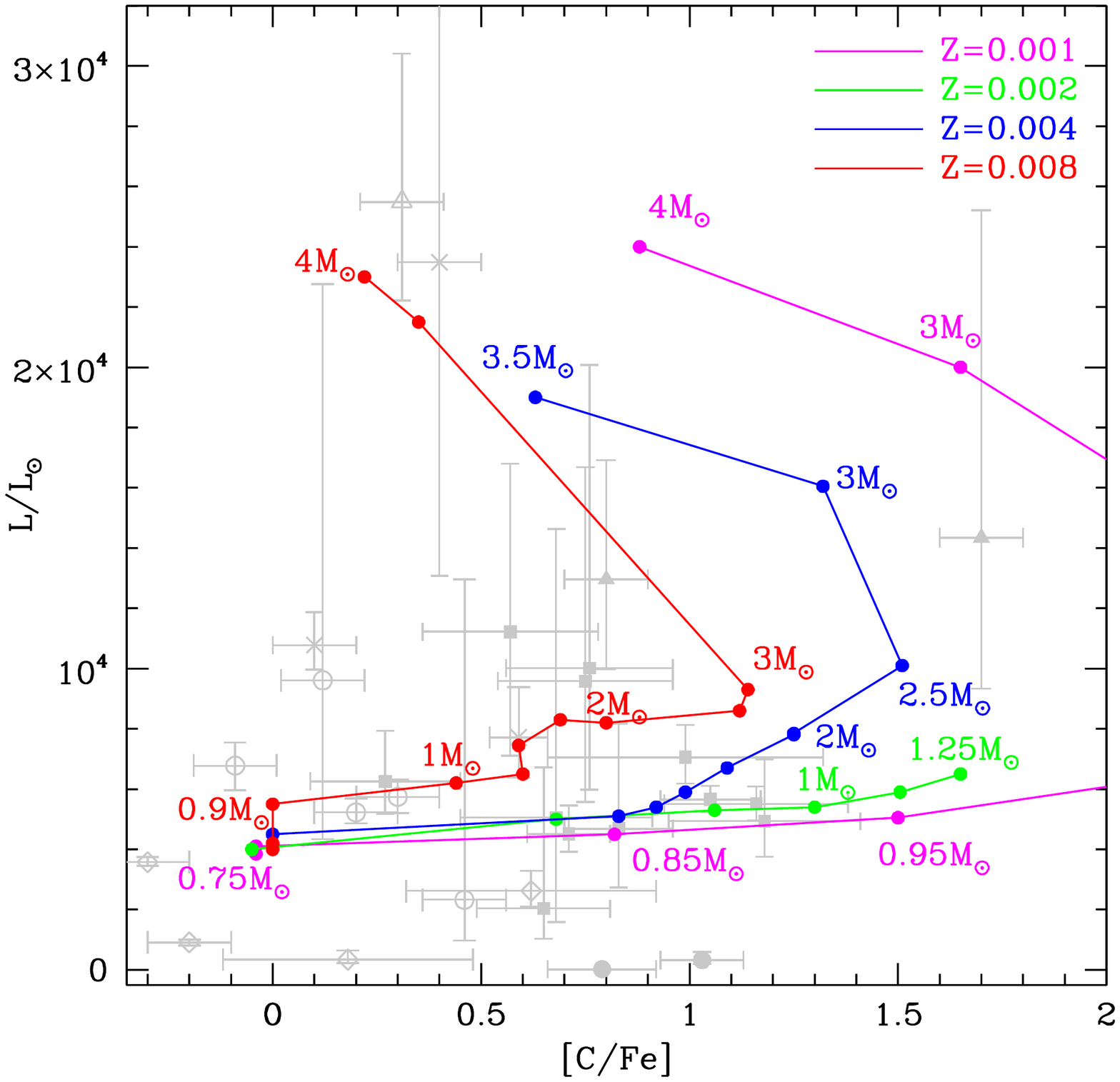}}
\end{minipage}
\vskip-60pt
\caption{Surface carbon and luminosity of the post-AGB stars in the sample presented in paper I, overimposed
to results from stellar evolution modelling, represented by the solid lines, that connect the expected
carbon and luminosity of model stars of different mass and same metallicity. The colour coding used for the
individual sources and for the lines is reported in the right panels. Stars showing s-process enrichment
are represented with full circles, whereas open points refer to sources with no sign of any s-process
enrichment. The meaning of the symbols used to represent the different sources is as follows: circles - low-mass stars, not experiencing any TDU episode; squares: stars that experienced TDU and not HBB; triangles: stars that experienced both TDU and HBB; asteriscks: stars that suffered an early loss of the external envelope.} 
\label{fclum}
\end{figure*}

The luminosity of post-AGB stars provides valuable information to identify the progenitors, as it is tightly connected with the core mass, thus with the initial mass of the star, that, in turn, allows the determination of the age and formation epoch. The core masses derived from the luminosity of post-AGB stars is more reliable than for AGBs, because in the latter stars the luminosity is subject to variations related to the occurrence of thermal pulses and to variability, whereas no significant variations in the overall energy flux is expected during the post-AGB phase.   

Among the chemical species involved in CNO cycling, the variation of the surface carbon is extremely sensitive to the initial mass of the star, as can be deduced from the values
reported in tab.~1, where we see that $[$C$/$Fe$]$ is a valuable indicator of both mass and metallicity of the stars. This follows from the arguments discussed in the previous section and partly shown in Fig.~\ref{fdup}, that the gradual accumulation of $^{12}$C in the external regions of the star 
depends of the number of TDU experienced, that grows with the mass of the star. The tight link between the surface carbon and the mass of the star holds even more for massive AGBs, as the most relevant effect of HBB is the destruction the surface carbon\footnote{The range of masses explored in tab. 1 is limited to $4~\rm{M}_{\odot}$ as there is no indication of the presence of the progeny of massive AGBs in the sample stars studied here; however the low surface carbon expected in  post-AGB sources descending from massive AGB stars must be considered for future investigations.}. 

The nitrogen content of post-AGB stars is less sensitive than carbon to the progenitor mass, because no significant changes in the surface $^{14}$N are expected to occur during the AGB phase of $\rm{M} \leq 3~\rm{M}_{\odot}$ stars.
Differences in the nitrogen content among stars in this mass domain are mostly connected to the different efficiency of mixing processes during the RGB phase.
A significant rise in the nitrogen content is found for stars experiencing HBB, descending from $\rm{M} > 3~\rm{M}_{\odot}$
progenitors. The $[$C$/$Fe$]$ and $[$N$/$Fe$]$ ranges reported in tab.~1 are similar and slightly above 1 dex, but $[$C$/$Fe$]$ is much more tightly connected to the mass of the star than $[$N$/$Fe$]$. A further reason why the carbon data are more useful than nitrogen for the present analysis is that $[$N$/$Fe$]$ is available for only $\sim 50\%$ of the sample stars.

Use of the oxygen abundance for the characterisation of the sources in the sample discussed here is not trivial, considering the behaviour of the surface $^{16}$O during the life of the star. Unlike carbon and nitrogen, no significant variation in the surface oxygen is expected during the RGB evolution. Oxygen depletion due to HBB is expected in metal-poor massive AGBs \citep{ventura13} and in extremely metal-poor $\rm{M} \geq 3~\rm{M}_{\odot}$ stars \citep{ventura21}; we do not expect to see any trace of this in the present context, as we will show that the sample stars from paper I descend from slightly sub-solar, $\rm{M} \leq 4~\rm{M}_{\odot}$ progenitors. Regarding low-mass stars ($\rm{M} \leq 3~\rm{M}_{\odot}$ progenitors), we see in tab. 1 that some oxygen enrichment is found in model stars within the $2-3~\rm{M}_{\odot}$ mass range. 
The expected increase in the surface $^{16}$O is within 
0.3 dex, thus much smaller than the variation of carbon and comparable with the typical errors associated to the derived abundances. A further issue connected to the use of oxygen for the characterisation of post-AGB stars is that the initial oxygen content is sensitive to the $\alpha-$enhancement of the gas from which the star formed; the signature of the $\alpha-$enhancement will unfortunately leave the imprinting on the post-AGB oxygen, because, unlike carbon and nitrogen, no significant changes that might erase the initial content are expected during the RGB and AGB phases.

On the basis of this discussion, we will consider the luminosity and the surface carbon of post-AGB stars as the most reliable witnesses of the previous AGB evolution of the individual sources, thus the most relevant factors to characterise the stars, in terms of mass and formation epoch of the progenitors, and of the efficiency of the different mechanisms which contributed to the modification of the surface chemical composition. This is the motivation for the choice of the carbon - luminosity plane as the privileged working tool for the present investigation, on which comparing the observations with results from stellar evolution modelling, from the infancy of the stars until the post-AGB, the phase that the sources in the sample stars are nowadays evolving through. The conclusions drawn from this analysis will be further strengthened or disregarded by looking for consistency, whenever possible, with the observed nitrogen and oxygen abundances, which are important to test the efficiency of physical mechanisms other than TDU, such as deep mixing during the RGB and the activation of HBB.

In Fig.~\ref{fclum} the surface carbon and luminosity of the sample stars are compared with those
derived from evolutionary calculations, represented by the coloured lines introduced in the previous
section. The sources are indicated with various symbols, each corresponding to the interpretation given in the present study, and (left panel) different colours, according to the metallicity.

The sample stars showing s-process enrichment, represented with filled symbols in Fig.~\ref{fclum}, are mostly composed by $1-1.5~{\rm M}_{\odot}$ stars of sub-solar metallicity, which experienced a series of TDU events, that favoured the surface carbon enrichment. These stars, indicated with squares in the figure, formed earlier than
2 Gyr ago. They are not expected to evolve at luminosities above $\sim 8000~{\rm L}_{\odot}$ (see tab. 1), which is 
consistent with the observations, although the error bars associated to the luminosities are so large
that no firm conclusions can be drawn. The surface nitrogen, available for about half of these sources,
is $0<[$N$/$Fe$]<0.6$ in all cases, consistent with the N enhancement expected during the RGB ascending; no further significant N enhancement is expected to
occur during the AGB phase of these stars. 

Two out of the stars belonging to the s-process enriched sub-sample, indicated with full triangles in 
Fig.~\ref{fclum}, are characterised by large luminosities, in the $10000-20000~{\rm L}_{\odot}$ range,
and surface nitrogen $[$N$/$Fe$]>1$. This is a clear signature of HBB, which favoured
the synthesis of large quantities of nitrogen at the base of the envelope, via CN cycling. 
The luminosity range and the measured carbon enrichment of the two stars suggest $3-3.5~\rm{M}_{\odot}$
progenitors, formed $\sim 300$ Myr ago. For both stars the oxygen abundances are not available; however,
as discussed previously, stars in this mass range are not expected to experience sufficiently 
strong HBB to lead to the depletion of the surface oxygen, independently of
metallicity \citep{ventura13}. 

Regarding the stars exhibiting no sign of s-process enrichment, indicated with open symbols in Fig.~\ref{fclum}, we interpret 5 of these sources as descending from low-mass progenitors of initial mass below $\sim 1~\rm{M}_{\odot}$. These are the oldest objects in the sample examined, formed definitively earlier than 5 Gyr ago\footnote{Unfortunately no firm conclusions on the age of these stars can be drawn, because as discussed in previous sections the masses deduced from the present analysis, based on the values reported in tab. 1, refer to the beginning of the AGB phase, whereas the age is connected with the initial mass of the star, which is sensitive to the unknown mass loss occurred during the RGB}. The surface carbon abundances of these stars are 
$[$C$/$Fe$]<0.5$. According to our understanding these stars started the AGB phase with an envelope mass
below $\sim 0.3-0.4~{\rm M}_{\odot}$, which was lost after a few TPs, typically 4 or 5, after the star
experienced at most a couple of TDU events, thus preventing significant s-process enrichment. Such TDU events
would be sufficient to raise the surface carbon after the depletion during the RGB phase, but the final
$[$C$/$Fe$]$ is not expected to grow in excess of the values detected for these stars. The nitrogen abundances
detected for the stars in this group are in agreement with the expectations regarding the N enrichment
occurring during the RGB phase, with the only exception of HD 161796, with $[$N$/$Fe$]=1.1$, which we discuss
separately.

The group of stars showing no trace of s-process enrichment also include one bright source (open triangle in
Fig.~\ref{fclum}), which we believe to have suffered HBB during the AGB evolution. The luminosity of this star is definitively above $20000~\rm{L}_{\odot}$, compatible with a $\sim 4-5~\rm{M}_{\odot}$ progenitor. The hypothesis of HBB is consistent with the large surface nitrogen, with $[$N$/$Fe$]$ slightly below 1. There is no trace of
carbon and s-process enrichment, which indicates poor or no effects from TDU.

\section{Post-AGB stars as tracers of the AGB evolution}
The characterisation of the individual sources addressed in the previous section confirms how
the observation of post-AGB stars is of extreme importance for obtaining valuable information
regarding the main aspects of the AGB evolution.
To further investigate the potentialities of this kind of analysis we focus on some open and highly
debated points related to the AGB phase, and show how present and future observations of post-AGB stars
might lead to significant steps forwards towards a more exhaustive comprehension of AGB stars.

\subsection{The luminosity threshold for carbon enrichment}
\label{tdulim}
While it is generally recognised that the formation of carbon stars is due to the occurrence of repeated TDU episodes \citep{iben82}, the efficiency of this mechanism and the evolutionary stage during the AGB evolution when the first TDU event is experienced are still debated. Results from observations of carbon stars in the Magellanic Clouds showed that use of the classic Schwarzschild neutrality criterion was not sufficient to reproduce the lower luminosity tail of carbon stars (e.g. Iben 1981), thus pushing modellers to introduce overshoot from the base of the convective envelope. Given the tight connection between core mass and the physical properties (primarily luminosity) of AGB stars, some studies were devoted to fix the core mass at which TDU begins during the AGB evolution \citep{martin93, marigo99}. 

The study of the threshold luminosity of carbon-enriched post-AGB stars turns out to be a more valuable, indirect indicator of the efficiency of TDU than the luminosities derived for carbon stars in the AGB phase.
Indeed in the latter case it is possible that the star is observed during an evolutionary phase following a thermal pulse, when the luminosity has not recovered the pre-TP value, thus preventing a reliable determination of the core mass of the star. In the post-AGB phase the luminosity is not exposed to meaningful variations, which allows a much more reliable indication of the core mass of the star. 

\begin{figure}
\begin{minipage}{0.48\textwidth}
\resizebox{1.\hsize}{!}{\includegraphics{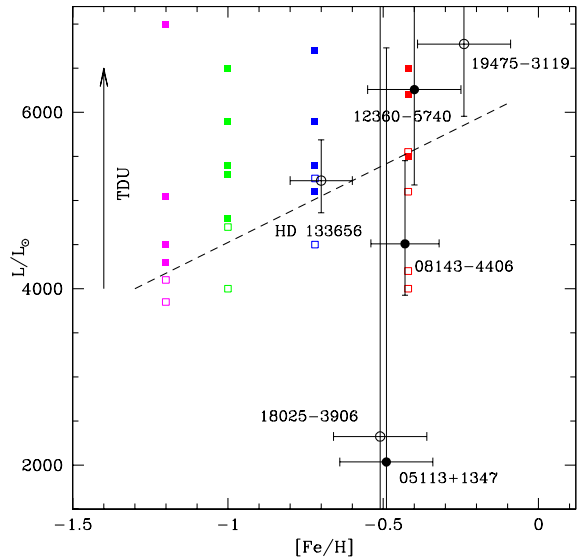}}
\end{minipage}
\vskip-60pt
\caption{The luminosity of the model stars reported in tab. 1, grouped according to the metallicity, reported on the abscissa. The same colour coding as in Fig.~\ref{fclum} is adopted. The dashed line separates the upper region, where
we expect to see some effects of the TDU, from the low-mass stars region, where no TDU during the AGB phase took place.
Black points refer to the sources discussed in section \ref{tdulim}, where full symbols indicate stars with some
s-process enrichment and open points refer to stars with no s-process.
}
\label{figZL}
\end{figure}

Regarding the models used in the present work, that similarly to the studies mentioned above 
assume overshoot from the borders of regions unstable to convection (see the beginning of section \ref{models}), the threshold mass and luminosity required to achieve carbon enrichment during the AGB phase can be deduced by the values reported in tab. 1. We find that the stars experiencing TDU descend from progenitors with masses at the beginning of the AGB in the $0.85-0.95~{\rm M}_{\odot}$ range, the threshold mass being larger the higher is the metallicity. This affects the minimum post-AGB luminosity below which we do not expect to detect carbon enriches
sources, which is in the $4500-5500~{\rm L}_{\odot}$ range, the lowest threshold luminosities corresponding
to the lowest metallicity (see tab. 1). This is outlined in Fig.~\ref{figZL}, that shows the luminosities reported in tab. 1, for stars of different mass and metallicity. The dashed line in the figure indicates the threshold luminosity below which we do not expect to detect the effects of TDU, as a function of the metallicity.

The sample presented in paper I include 6 sources, which according to our understanding, based on the observed surface chemistry and the derived luminosities, descend from stars that either just failed to achieve any carbon and s-process enrichment, or experienced only a very few TDU, sufficient to slightly raise the surface carbon with respect to the abundance at the end of the RGB phase. These stars are the ideal target to study the efficiency of 
TDU and the evolutionary stage during the AGB phase of low-mass stars, when we expect this mechanism to begin. In the first group we find HD 133656, IRAS 18025-3906, IRAS 19475+3119, whereas in the second sub-sample we find IRAS 05113+1347, IRAS 08143-4406 and IRAS 12360-5740. The metallicity and luminosity of these stars, with the relative error bars, are reported in Fig.~\ref{figZL}, where the luminosities of the stars of different mass and
metallicity included in tab. 1 are also shown, and indicated with full squares.

According to the evolutionary models used in the present analysis we expect that HD 133656 IRAS and 18025-3906 have luminosities slightly below $5000~\rm{L}_{\odot}$, whereas for IRAS 19475+3119, which is the highest metallicity source in the sample, we expect $\rm{L} \sim 6000~\rm{L}_{\odot}$. Regarding the latter three stars, all characterised by a slightly sub-solar metallicity, we expect that the luminosities are in the $5500-6000~\rm{L}_{\odot}$ range. These values are compatible, within the observational errors, with the luminosities given for the above sources, but as clear from Fig.~\ref{figZL} the error bars are too large to draw firm conclusions; this holds in particular for IRAS 18025-3906 and IRAS 05113+1347, for both of which the luminosity ranges from values non consistent with any previous AGB evolution to luminosities close to those expected for stars that experienced HBB. The study of these sources offers a unique opportunity to draw information on the conditions under which low-mass stars experience TDU, thus on the minimum luminosity at which we expect to observed carbon stars in galaxies. A tighter determination of the luminosity of these 6 stars is needed to test the validity of the modelling used here, or to understand the corrections in the treatment of convective borders required to reproduce the observations 

\subsection{Hot bottom burning}
The possibility that AGB stars experience HBB dates back to the 80's \citep{renzini81} and was further investigated by \citet{blocker91}, who first proposed that the stars experiencing HBB deviate significantly from the classic core mass - luminosity relationship derived by \citet{paczynski}. The study by \citet{ventura05} stressed the role played by convection modelling on the determination of the temperature at the base of the convective envelope, thus on the strength of HBB: in particular, the FST treatment used here was demonstrated to lead to much stronger HBB conditions with respect to models based on the classic MLT description. As discussed in section \ref{chem}, HBB is experienced by $M \geq 3-3.5~{\rm M}_{\odot}$, whereas stars formed with mass below this minimum threshold do not experience any HBB, thus their surface chemical composition may be altered by TDU only during the AGB phase.
 
As discussed in section \ref{chem}, the occurrence of HBB leaves important signatures in the surface chemical composition of the stars, the most impressive effect being the large nitrogen enhancement: in the example shown in Fig.~\ref{fhbb} the final N, with which the star is expected to enter the post-AGB phase, is $\sim 20$ times higher than the initial content. For the stars that experience HBB, the knowledge of the surface carbon when
evolving through the AGB is a valuable indicator of the relative efficiency of TDU and HBB in the modification of the surface chemistry, considering their opposite effect on the surface $^{12}$C, which is destroyed by HBB and
produced by TDU. Theoretically the surface oxygen abundance of post-AGB stars might also provide important information on the strenght of the HBB experienced, as significant reduction of the surface oxygen is expected in stars experiencing strong HBB. However, this cannot be tested in the present study, because the depletion of oxygen is expected only in metal-poor environments \citep{ventura13, flavia18a}, whereas the present sample is mostly composed by slightly sub-solar metallicity stars.

\begin{figure*}
\begin{minipage}{0.32\textwidth}
\resizebox{1.\hsize}{!}{\includegraphics{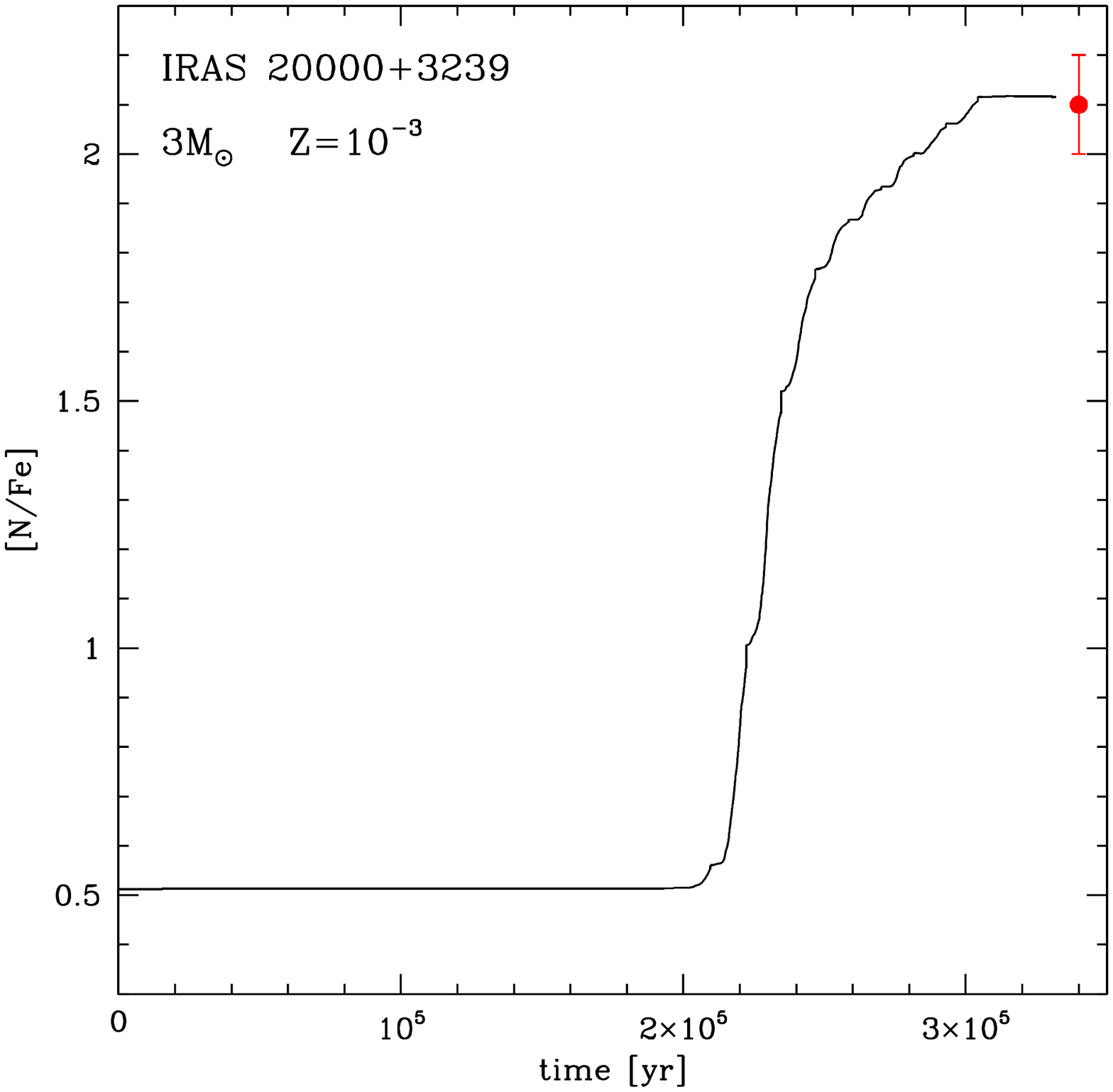}}
\end{minipage}
\begin{minipage}{0.32\textwidth}
\resizebox{1.\hsize}{!}{\includegraphics{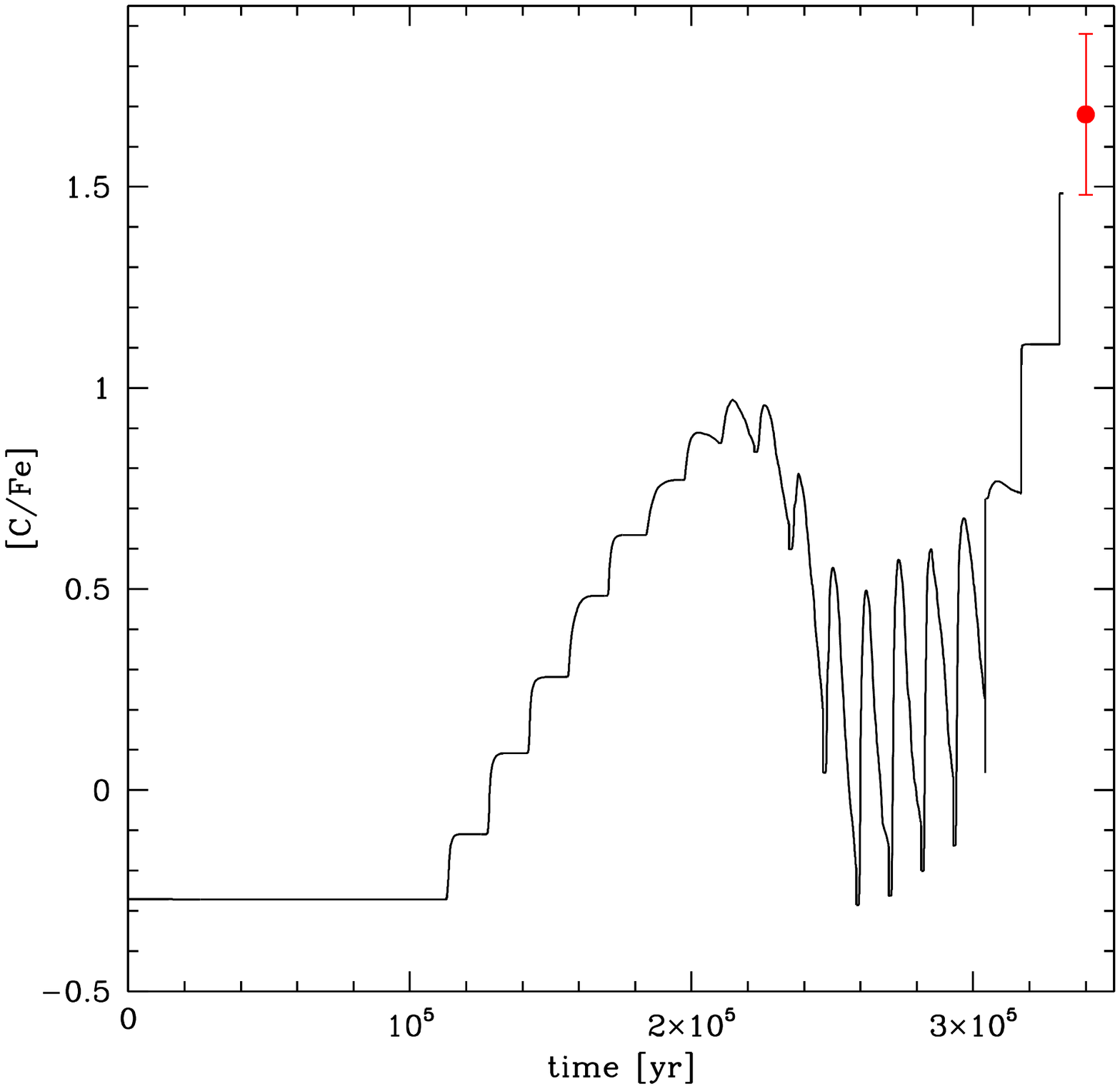}}
\end{minipage}
\begin{minipage}{0.32\textwidth}
\resizebox{1.\hsize}{!}{\includegraphics{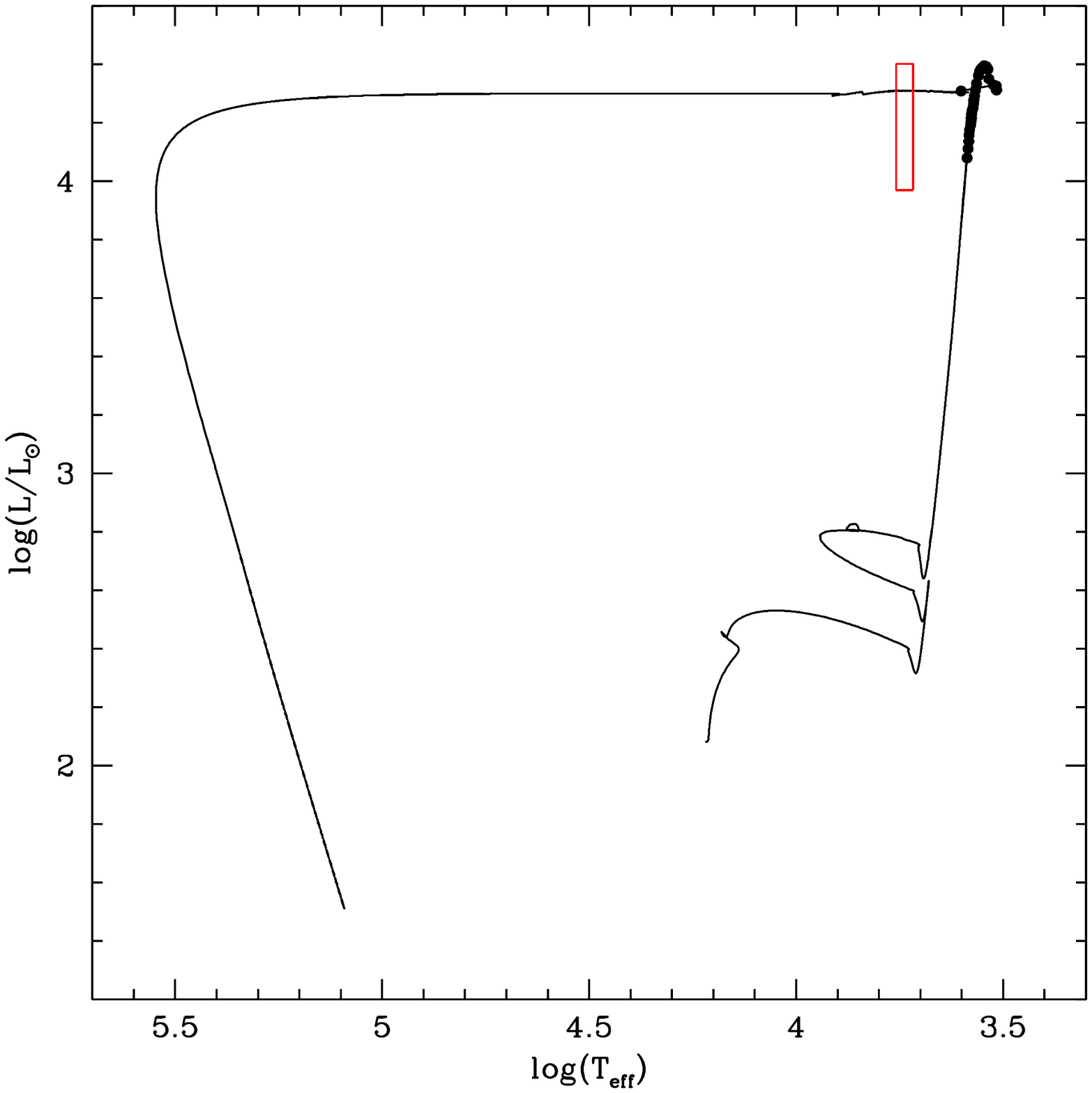}}
\end{minipage}
\vskip-40pt
\caption{AGB variation of the surface nitrogen (left panel) and carbon (middle panel) of a $3~{\rm M}_{\odot}$
model star of metallicity $Z=10^{-3}$, as a function of the AGB lifetime. Red points with the error bars
refer to the measure abundances of the source IRAS 20000+3239. The right panel reports the evolutionary
track of the same model star on the HR diagram and the effective temperature and luminosity of 
IRAS 20000+3239, with the associated error box.}
\label{f20000}
\end{figure*}

In the previous section we have identified 3 sources in the sample, namely IRAS 02229+6209, IRAS 20000+3239 and HR 6144, as bright stars, enriched in nitrogen, which experienced HBB during the AGB phase. IRAS 02229+6209 and IRAS 20000+3239 exhibit significant s-process enhancement, whereas for HR 6144 the situation is more tricky,
with $[$s$/$Fe$]=0.2 \pm 0.2$, thus a modest s-process enrichment.

In Fig.~\ref{f20000} we show the time evolution of the surface chemical composition of a $3~\rm{M}_{\odot}$ star with metallicity $Z=10^{-3}$, which we believe to be a possible progenitor of IRAS 20000+3239. The effects of HBB are seen in the rise of the surface nitrogen (left panel), which grows by a factor $\sim 40$ with respect to the N abundance at the beginning of the AGB phase\footnote{In the figure, which refers to the AGB phase, the initial surface $[$N/Fe$]$ is $\sim 0.5$. The evolution was calculated with a solar-scaled nitrogen: the super-solar $[$N/Fe$]$ at the beginning of the AGB phase is due to the combined effects of the first and second dredge-up events.}. The behaviour of the surface carbon is a key indicator of the combined effects of TDU + HBB: during the first part of the AGB evolution the surface $^{12}$C increases owing to the effects of TDU, whereas after $\sim 2\times 10^5$ yr we note the effects of HBB, in the fast drop of the surface carbon during the intepulse phases. During the very final AGB phases, after HBB was turned off by the loss of the envelope, the effects of TDU take over, and the surface $^{12}$C rises again, until reaching the final value of $[$C/Fe$]\sim 1.5$, with which the star is expected to evolve to the post-AGB phase. 

The final values of the surface C and N are consistent with the abundances given for IRAS 20000+3239. The evolutionary track for this star (right panel) is also consistent with the observations, with the expected luminosity during the post-AGB phase, $\rm{L} \sim 20000~\rm{L}_{\odot}$, well within the observational uncertainties. A further support to the interpretation given for IRAS 20000+3239 is the detection of lithium in the atmosphere. As discussed in section \ref{chem}, lithium can be observed in post-AGB stars only if the HBB experienced is not strong, in such a way that $^3$He is not totally consumed before the start of the post-AGB phase: this is the case for the $3~\rm{M}_{\odot}$ model star proposed as progenitor for
IRAS 20000+3239, because the HBB experienced is not particularly strong, with temperatures at the base
of the envelope below $70$MK. IRAS 20000+3239 is the most metal-poor star in the sample: massive AGB stars of similar metallicity are exposed to strong HBB, which triggers the depletion of the surface oxygen. Unfortunately this possibility cannot be tested here, not only because the surface oxygen is not available for this source, but also because the progenitor
mass is lower than the $\sim 5~{\rm M}_{\odot}$ limit required to ignite strong HBB, that for the metallicity
of IRAS 20000+3239 leads to the depletion of the surface $^{16}$O \citep{flavia18a}.

The discussion for IRAS 20000+3239 can be similarly extended to IRAS 02229+6209, which also exhibits carbon and 
nitrogen enhancement, with $[$C/Fe$]= 0.8$ and $[$N/Fe$]= 1.2$. For this source we also expect a progenitor of mass just above the minimum threshold required to activate HBB, $\sim 3.5~M_{\odot}$\footnote{The minimum threshold mass to reach HBB conditions slightly depends on metallicity, and changes from $\sim 3~{\rm M}_{\odot}$, for $Z=10^{-3}$, to $\sim 3.5~{\rm M}_{\odot}$, for $Z=8\times 10^{-3}$}, that formed $\sim 250$ Myr ago.
Both TDU and HBB leave important signatures in the surface chemical composition of these stars. In the carbon - luminosity plane shown in Fig.~\ref{fclum}, this source is found in a different position with respect to IRAS 20000+3239, despite the similarity in the progenitor mass: as discussed in section \ref{chem}, this is due to the higher metallicity of IRAS 02229+6209, which, for a given increase in the surface abundances of carbon or nitrogen, makes the corresponding increase in $[$C/Fe$]$ or $[$N/Fe$]$ smaller. According to our modelling the luminosity of IRAS 02229+6209 is $\sim 16000-18000~\rm{L}_{\odot}$, which is consistent with the $10000-17000~\rm{L}_{\odot}$ range given by the observations: future confirmation of luminosity values close to the lower limit within the observational error would be at odds with the current interpretations, as no HBB is expected to take place in stars then enter the post-AGB phase with ${\rm L} < 15000~\rm{L}_{\odot}$. 

Among the 3 stars that experienced HBB during the AGB phase, HR 6144 is the brightest,
with the luminosity estimated in the $23000-30000~\rm{L}_{\odot}$ range. The surface carbon of this source is significantly smaller than IRAS 20000+3239 and IRAS 02229+6209, and there is no trace of s-process enrichment. These results are consistent with the hypothesis of a higher mass progenitor, in which the effects of HBB were dominating over TDU during the AGB lifetime, which is the reason for the smaller surface carbon detected and the lack of significant s-process enrichment. In the comparison with IRAS 02229+6209, a source with similar metallicity, the $[$N/Fe$]$ is smaller: this further reinforces the higher mass progenitor hypothesis, as in this case nitrogen would be produced via HBB by the $^{12}$C originally present in the star, whereas for stars like IRAS 02229+6209 there is an additional contribution from the burning of primary carbon, dredged-up from the helium-burning shell. A further difference with respect to IRAS 02229+6209, which is also consistent with the possibility that HR 6144 descends from a higher mass progenitor, is the lack of lithium: this is because the HBB experienced by HR 6144 is stronger, with temperatures at the base of the envelope of the order of 90 MK, thus the lithium-rich phase is concluded before the end of the AGB evolution, as the consumption of $^{3}$He is extremely fast; on the qualitative side, this case, as far as the evolution of the surface lithium is concerned, is similar to the one shown in Fig.~\ref{fhbb}.

Finally, the interpretation proposed here is consistent with the large luminosity of HR 6144, compatible with progenitors in the $4-5~\rm{M}_{\odot}$ range. Based on the above arguments, we conclude that HR 6144 formed 100-200 Myr ago, thus being the youngest object in the sample presented in paper I. 

Further confirmation for the origin of this source might come from the determination of the sodium abundance,
as the temperatures mentioned above are sufficient for the activation of the Ne-Na nucleosynthesis, which
should have favoured the sodium synthesis, with $[$Na$/$Fe$] = 0.8-1$. Some aluminium production, due to
proton captures by the heavy isotopes of magnesium $^{25}$Mg and $^{26}$Mg is expected too, but the 
the overall Al increase is within +0.3 dex, thus not easy to be confirmed by the observations. The behaviour of sodium and aluminium expected for this source is shown in Fig.~\ref{fhbb}. For the
metallicity of HR 6144 no depletion of $^{24}$Mg, which would favour a much larger Al increase, is expected
\citep{flavia18a}. An additional check of the reliability of the interpretation given for this source
is the determination of the $^{12}$C/$^{13}$C ratio, which is expected to be in the 3-20 range. 
Carbon ratios of the order of $3-4$ correspond to the equilbrium values of the proton capture 
nucleosynthesis and would indicate a pure HBB effect; on the other hand $^{12}$C/$^{13}$C of the 
order of 10 or more would witness a contribution from TDU.

Overall, the results from the study of these 3 sources are extremely interesting for the 
research on massive AGBs, experiencing HBB. Indeed two out of the 3 stars descend from 
progenitors whose core mass and temperatures at the base of the envelope were just above 
the thresholds required to activate HBB, whereas the latter source discussed, HR 6144, 
descends from a higher mass progenitor, that experienced stronger HBB. Therefore the
analysis of these stars offer an interesting opportunity to test the predictions regarding 
the occurrence and the strength of HBB during the AGB phase, which, in turn, is connected 
with the description of the convective instability \citep{ventura05}. On this regard, 
considering that all these stars are characterised by a largely uncertain luminosity, 
a more robust determination of the overall energy flux is urgently needed for their use 
as a kay indicator of the HBB mechanism during the AGB lifetime 

Based on the interpretation reached on the basis if the present day data, we deduce that strong HBB conditions, similar to those experienced by HR 6144, quench TDU, such that the surface chemistry is mainly altered by HBB, with no effects from TDU. Further studies are required to add more robustness to this conclusion, which will be confirmed or disregarded according to whether future observations of post-AGB stars will confirm that all the brightest sources populate the same region of the carbon - luminosity plane as HR 6144, or if some bright post-AGB are s-process enriched and are located on the upper, right side of the plane shown in Fig.~\ref{fclum}.

\subsection{A signature of deep mixing during the RGB?}

\begin{figure*}
\begin{minipage}{0.48\textwidth}
\resizebox{1.\hsize}{!}{\includegraphics{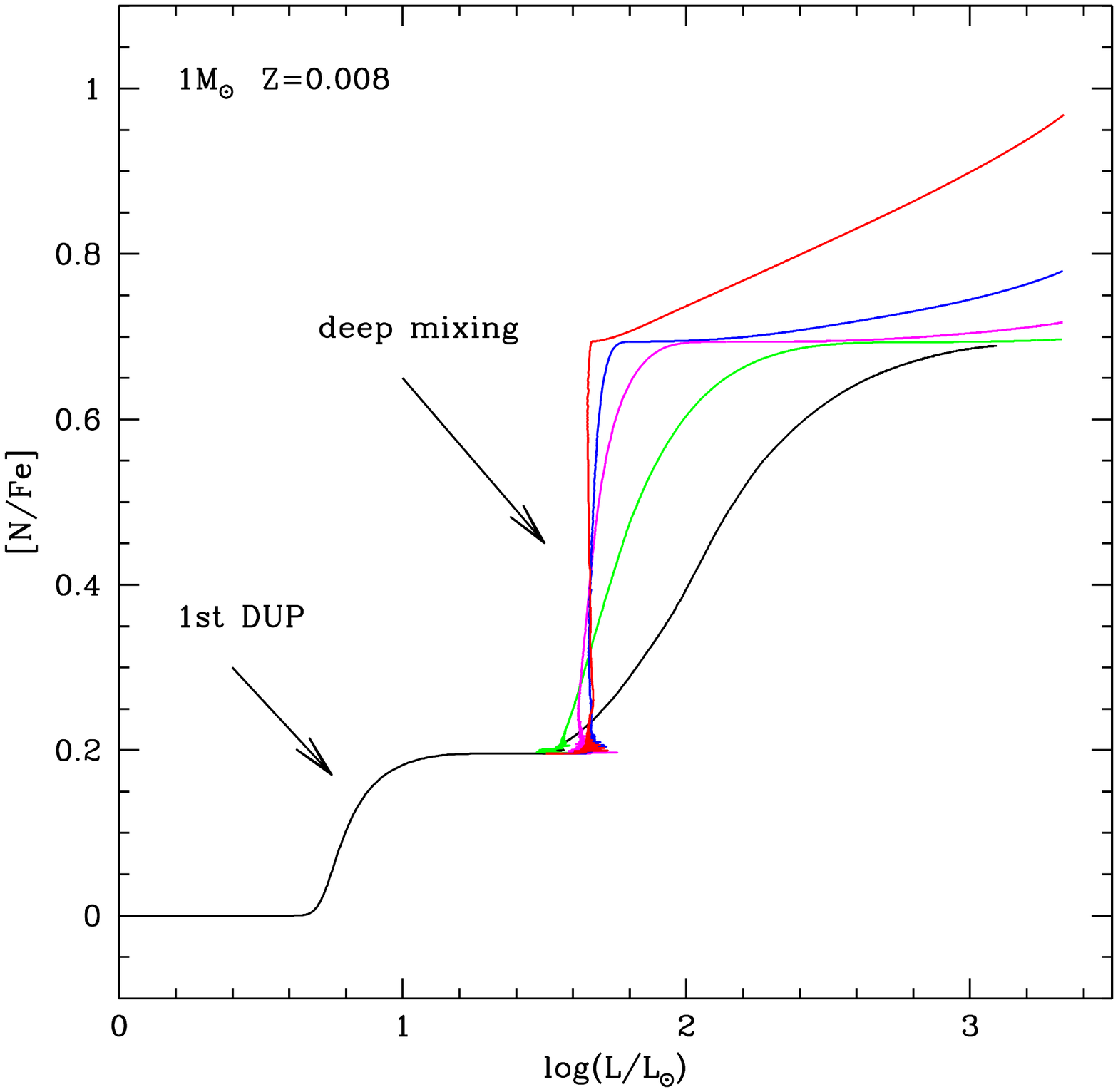}}
\end{minipage}
\begin{minipage}{0.48\textwidth}
\resizebox{1.\hsize}{!}{\includegraphics{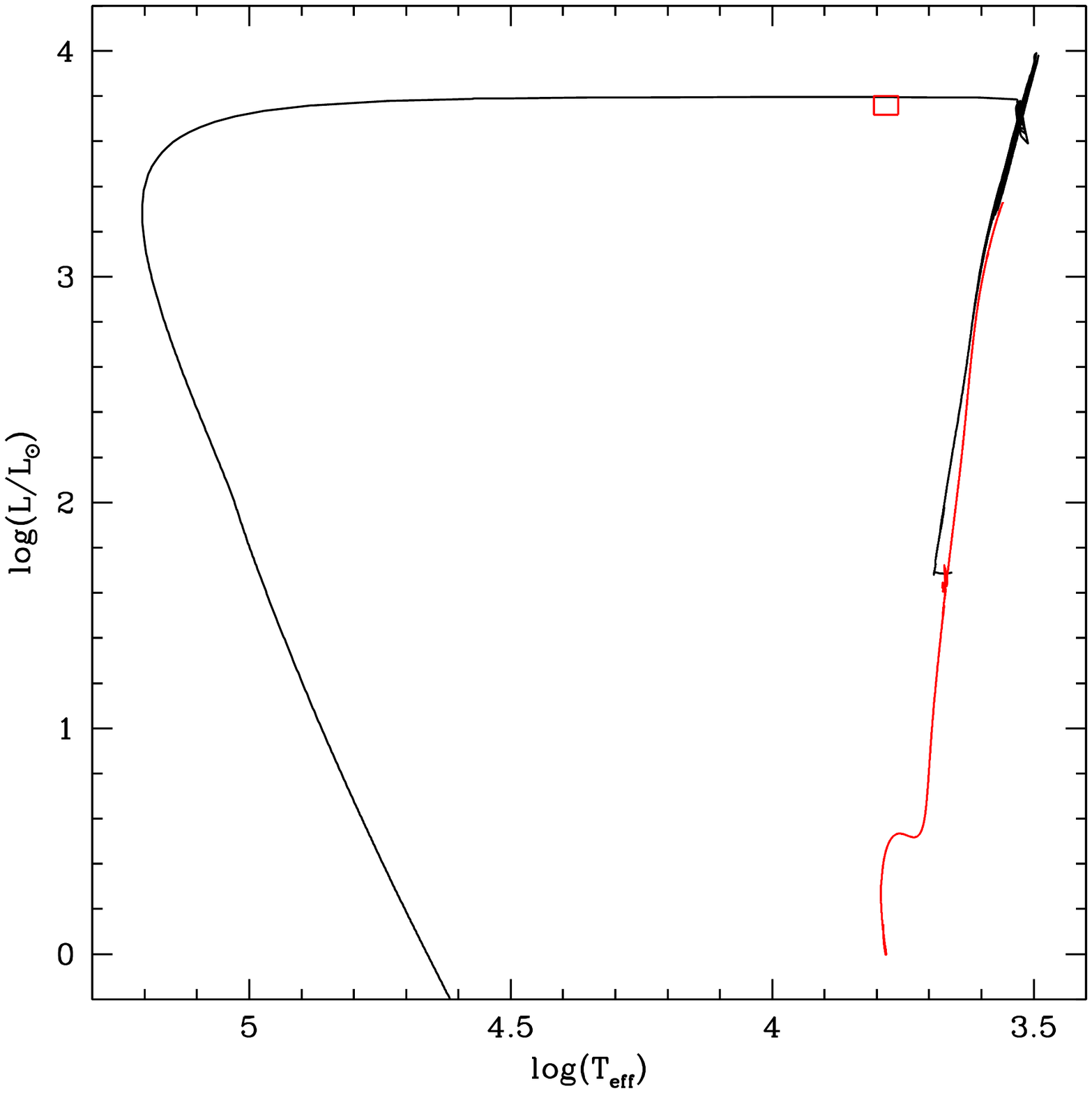}}
\end{minipage}
\vskip-60pt
\caption{Left: variation of the surface nitrogen of a $1~{\rm M}_{\odot}$ model star during the RGB phase,
as a function of the luminosity. The colour-coding refers to the efficiency of the assumed extra-mixing
from the base of the convective envelope, that begins after the stars evolves through the luminosity bump. Right: the evolutionary track of the $1~{\rm M}_{\odot}$ model star, split into the pre-helium flash (red line) and 
post helium-burning evolution (black); the red rectangle indicates the derived effective temperature and
luminosity of HD 161796, with the observational uncertainties.}
\label{fdmix}
\end{figure*}

In section \ref{interp} we identified 5 of the sources in the sample stars analysed in the 
present work as the descendants of low-mass progenitors, that lost the surface envelope after having experienced only a few TPs, with no significant effects of TDU. One of these stars, namely HD 161796, is characterised by
an extremely large surface nitrogen, $[$N/Fe$]= 1.1$. Large N enhancements are commonly associated to the
action of HBB, as discussed in the previous sections. We rule out this possibility for this source, as the post-AGB luminosities of the stars that experienced HBB are definitively above $\sim 15000~\rm{L}_{\odot}$,
far in excess of the $5000-6300~\rm{L}_{\odot}$ values indicated by the observations. 

A possible explanation for the unusually large nitrogen derived for this source is extremely deep mixing during the RGB ascending. This possibility was explored e.g. by D'Antona \& Ventura (2007), in relation to the presence of oxygen-poor giants in globular clusters. 
The effects of deep RGB mixing of various depths on the surface nitrogen is shown in the left panel of Fig.~\ref{fdmix}, where it is reported the variation of the surface N as a function of the luminosity of the star, starting from the post-MS phase until the ignition of the helium flash. It is clear in the figure the $\sim 0.2$ dex increase in the surface N, due to the effects of FDU, then the further rise in $[$N/Fe$]$, due to deep mixing. The surface nitrogen abundance at the ignition of the helium flash is approximately the same with which the star evolves through the post-AGB phase, since no significant variation in the surface N is expected during the AGB evolution of low-mass stars.

Our interpretation for HD 161796 is that it started the AGB phase with a mass in the $1-1.1~\rm{M}_{\odot}$
range. Assuming a $\sim 0.1~{\rm M}_{\odot}$ mass loss during the RGB, this corresponds to age $4-5$ Gyr.
Lower masses are ruled out, because the star must have experienced one or 2 TDU events before entering the 
post-AGB phase, otherwise the surface carbon left behind by deep mixing would be $[$C/Fe$]<0$ \citep{dantona07}, 
whereas the 
observed value is $[$C/Fe$]= 0.3$. We disregard higher mass progenitors, as in that case the star should
show up a larger carbon (see tab. 1) and significant s-process enrichment, which is not observed.

The evolutionary track of the proposed model is shown in the right panel of Fig.~\ref{fdmix}, where we see consistency with the values of effective temperature and the upper limit of the luminosity range derived by the observations.
A more robust estimate of the luminosity of this source is required before the explanation proposed here can be considered fully reliable.

\subsection{Stars that failed the third dredge-up}
\citet{devika17} discovered a bright star in the Small Magellanic Cloud that showed up no carbon and
s-process enrichment, thus did not experience any TDU. This finding was in tension with standard AGB evolution modelling, because the stars entering the post-AGB phase with the luminosities found in the above paper, 
around $8000~\rm{L}_{\odot}$, which correspond to $1.5-2~\rm{M}_{\odot}$ progenitors, are expected to exhibit significant carbon and s-process enrichment, as can be seen in the values reported in tab. 1. The \citet{devika17} discovery gave origin to a debate regarding physical mechanisms potentially able to affect the efficiency of TDU during the AGB evolution. Post-AGB stars are the ideal laboratory for this study, because the observed carbon is a key indicator of the efficiency of TDU during the whole AGB phase, and furthermore the derivation of the luminosity allows the determination of the core mass, hence of the mass of the progenitor.

In the sample discussed here we find 3 objects similar to the one studied by \citet{devika17}, which show no (or poor) carbon enhancement, against the results from AGB modelling: IRAS 23304+6147, HD 112374 and SAO 239853. These sources have luminosities definitively above $6500~\rm{L}_{\odot}$, therefore, according to the results reported in tab. 1,
are expected to be strongly enriched in carbon. The occurrence of strong HBB during the previous AGB phase might explain these data, but this interpretation could potentially hold only for SAO239853, since the luminosity range given for both IRAS 23304+6147 and HD 112374 are below the minimum threshold for the stars that experienced HBB, $\rm{L} \sim 15000~\rm{L}_{\odot}$.

A possible way to characterise post-AGB stars sharing properties similar to the one discovered by \citet{devika17} is considering models in which TDU is artificially inhibited, and identify the progenitor mass which is expected to enter the post-AGB phase with the core mass corresponding to the luminosity observed.  This approach, when compared with the results based on the models so far discussed, would lead to lower mass progenitors, as the core mass of a model star of a given initial mass grows bigger if the TDU is weaker. An alternative possibility is to consider model stars that evolve until reaching the core mass corresponding to the post-AGB luminosity and that experienced the number of TDU events required to obtain the observed carbon, then to assume that a fast loss of the external envelope halted  further growth of the core mass and increase in the surface carbon. 

While we cannot rule out the action of physical mechanisms that in some specific cases reduce the efficiency of TDU, we do not consider this possibility in the present analysis, because evolutionary results with no extra-mixing from the base of the envelope showed inconsistency with the observations, as the final surface carbon was still much larger than observed. We therefore explore the second possibility, by trying to identify the progenitors of the three sources considered based on their AGB evolution, looking for consistency between the values of carbon and luminosity derived from the observations with those attained by the model stars during specific evolutionary stages during the AGB lifetime, when we assume that the envelope is lost. The three sources discussed will be discussed separately.

\subsubsection{IRAS 23304+6147}
This source, with sub-solar metalliticy $[$Fe$/$H$] \sim -0.8$, is s-process enriched, which witnesses the action of some TDU episodes. The luminosity is in the $6500-10000~\rm{L}_{\odot}$ range, which is the luminosity with which stars of initial masses $1.5~\rm{M}_{\odot} < M < 2~\rm{M}_{\odot}$ and metallicity similar to that of IRAS 23304+6147 enter the post-AGB, at the end of their AGB lifetime. However, these stars are expected to reach
carbon abundances $[$C/Fe$] > 1$ (see tab. 1), whereas the observations indicate $[$C/Fe$] = 0.6$. 

\begin{figure}
\begin{minipage}{0.48\textwidth}
\resizebox{1.\hsize}{!}{\includegraphics{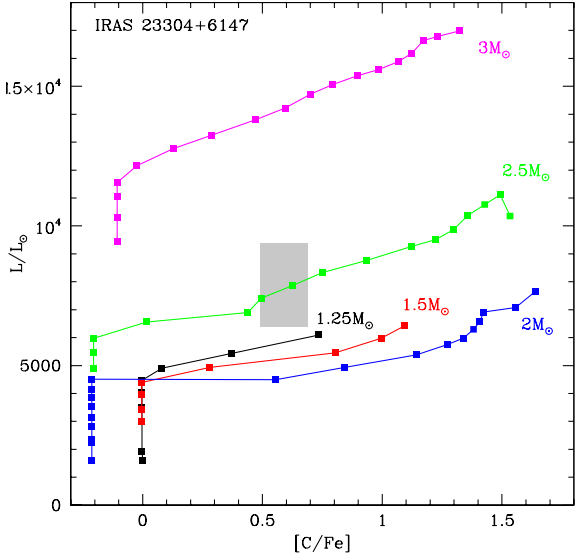}}
\end{minipage}
\vskip-60pt
\caption{The AGB variation of the surface carbon and luminosity of $1.25-3\rm{M}_{\odot}$ model stars of metallicity $Z=4\times 10^{-3}$. The different points correspond to the inter-pulse quantities.
The grey box indicates the surface carbon and luminosity of IRAS 23304+6147 given in paper I, with
the uncertainties.
}
\label{f23304}
\end{figure}

Fig.~\ref{f23304} shows the tracks of the stars with metallicity similar to
IRAS 23304+6147, with mass at the beginning of the AGB phase spanning the
$1.25~\rm{M}_{\odot} < M < 3~\rm{M}_{\odot}$ range. Lower mass are excluded, as they evolve at luminosities lower than observed, whereas higher mass are ruled out as they are too bright. The various points along each track correspond to different inter-pulse phases. 

Under the hypothesis that the envelope of the star is rapidly lost at a given evolutionary stage during the AGB, the comparison with the observed carbon and derived luminosities (grey box in Fig.~\ref{f23304}) allows the identification of the progenitor of the source discussed. Indeed the possibility that this source descends from $M < 2~\rm{M}_{\odot}$ progenitors can be disregarded, as these stars reach the derived luminosity only during the very final AGB phases, after significant carbon enrichment, far in excess of the observed value, occurred. $M > 2.5~\rm{M}_{\odot}$ progenitors can be also ruled out, as they evolve too bright since the very early AGB phases, thus their luminosity would be significantly higher than $\sim 10000~\rm{L}_{\odot}$, the largest value within the observational uncertainties.

\subsubsection{HD 112374}
This metal-poor source populates a region in the plane shown in Fig.~\ref{fclum} not covered by the
evolutionary tracks: for the luminosity of HD 112374, around $10000~\rm{L}_{\odot}$, the theoretical lines indicate large carbon enrichments, at odds with the surface carbon derived for this source, which is approximately solar-scaled. The surface carbon is consistent with low mass progenitors, with initial mass below $\sim 1~\rm{M}_{\odot}$, which on the other hand do not reach the luminosity of HD 112374. The possibility that the progenitor is a massive AGB star that experienced HBB can be disregarded too, not only because the luminosity is below the minimum threshold at which we expect that the stars undergoing HBB enter the post-AGB phase, but also because the observed nitrogen, $[$N$/$Fe$] = 0.5$, is not sufficiently large to be consistent with any effect from HBB, which require $[$N$/$Fe$] > 1$. 

Following the same arguments used to characterise IRAS 23304+6147, we attempt to identify the possible progenitor
of this source among the model stars that evolve at luminosities $\sim 10000~\rm{L}_{\odot}$ during the first
part of the AGB phase, after a few TPs, with no TDU experienced. In this case we use $Z=10^{-3}$ models as
$[$Fe$/$H$] = -1.2$ for HD 112374. This criterion allows us to deduce that the mass of the progenitor of
this source is a $2.5-3~\rm{M}_{\odot}$ range, formed $300-500$ Myr ago.

\begin{figure*}
\begin{minipage}{0.32\textwidth}
\resizebox{1.\hsize}{!}{\includegraphics{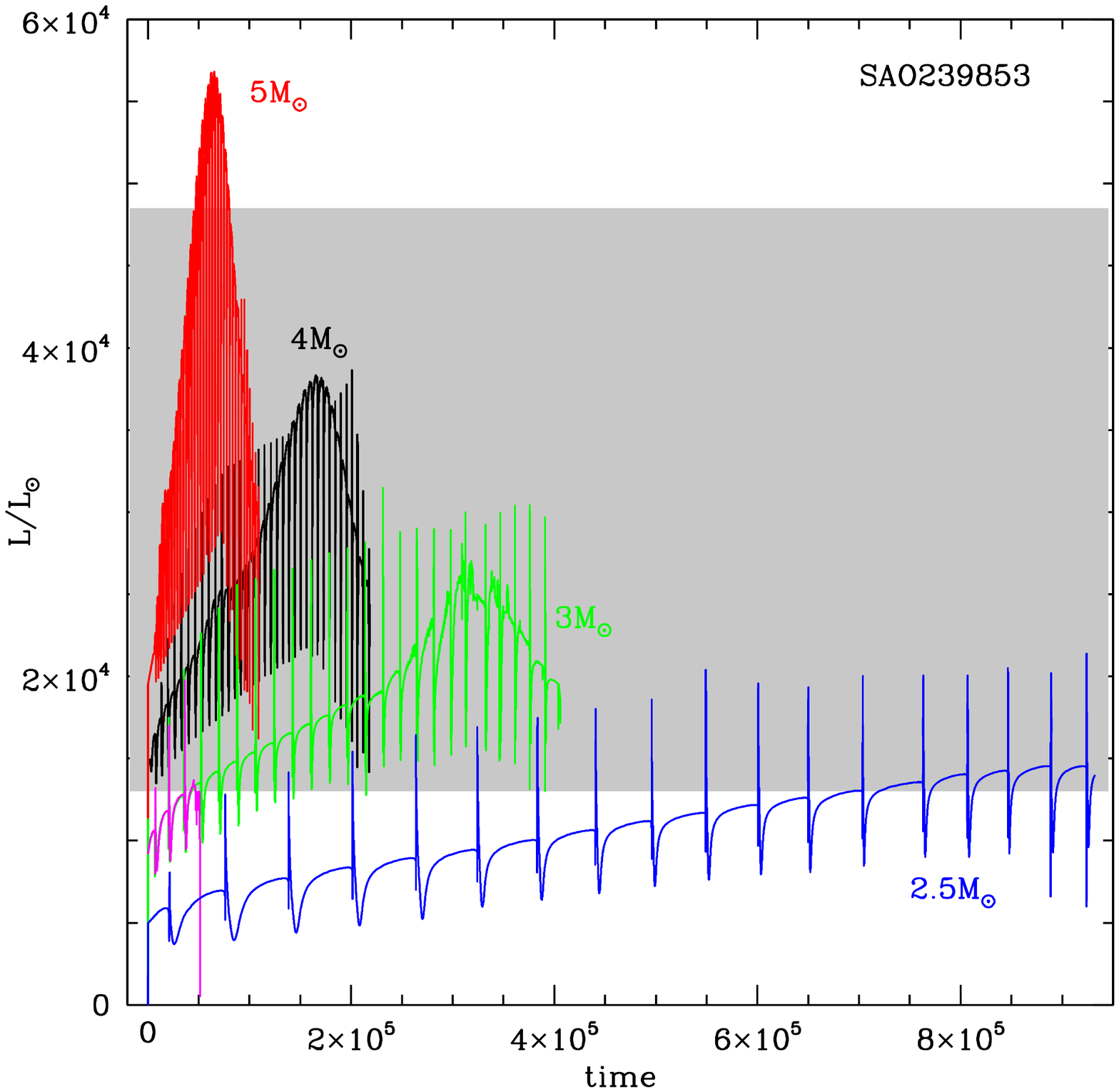}}
\end{minipage}
\begin{minipage}{0.32\textwidth}
\resizebox{1.\hsize}{!}{\includegraphics{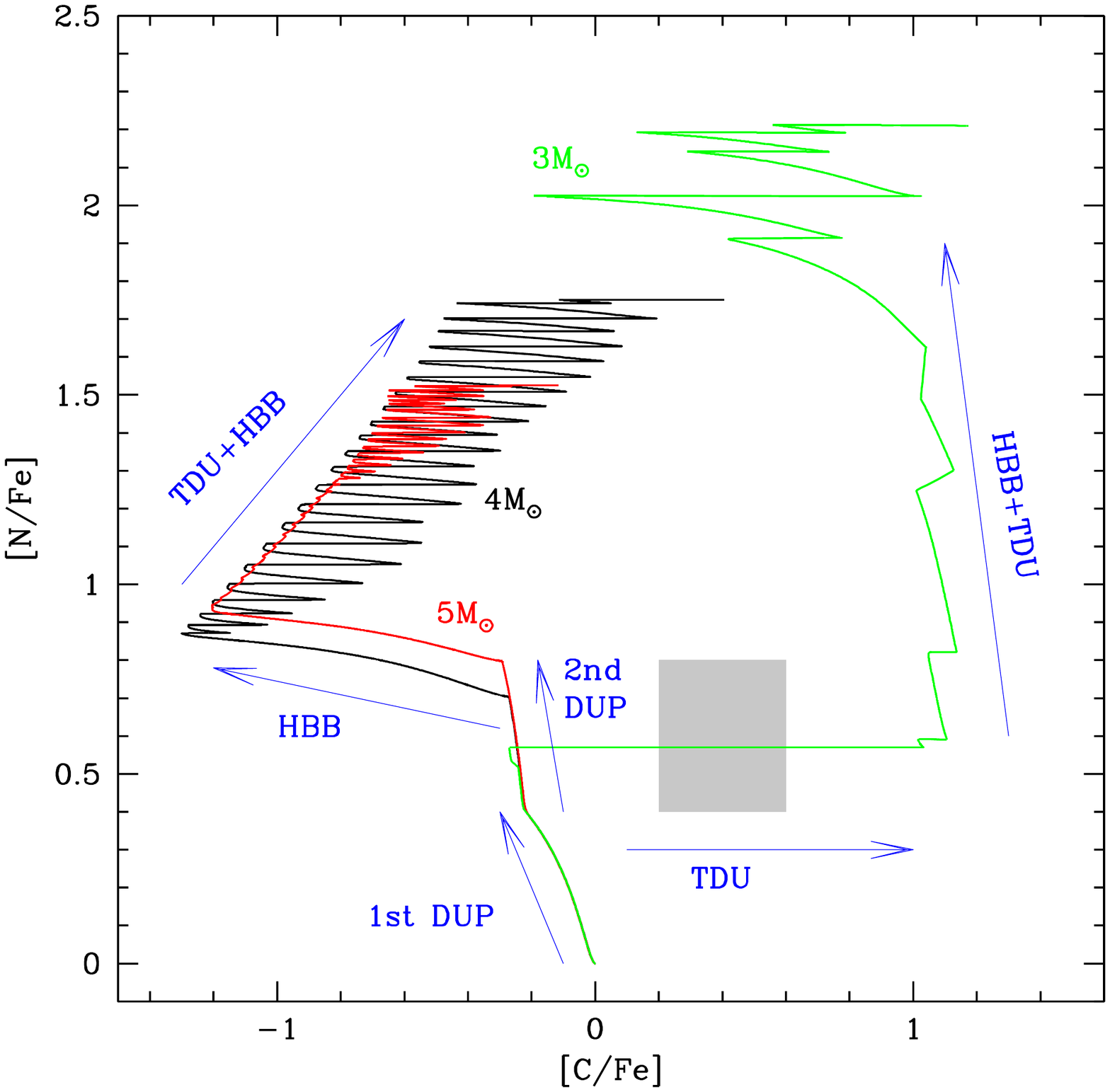}}
\end{minipage}
\begin{minipage}{0.32\textwidth}
\resizebox{1.\hsize}{!}{\includegraphics{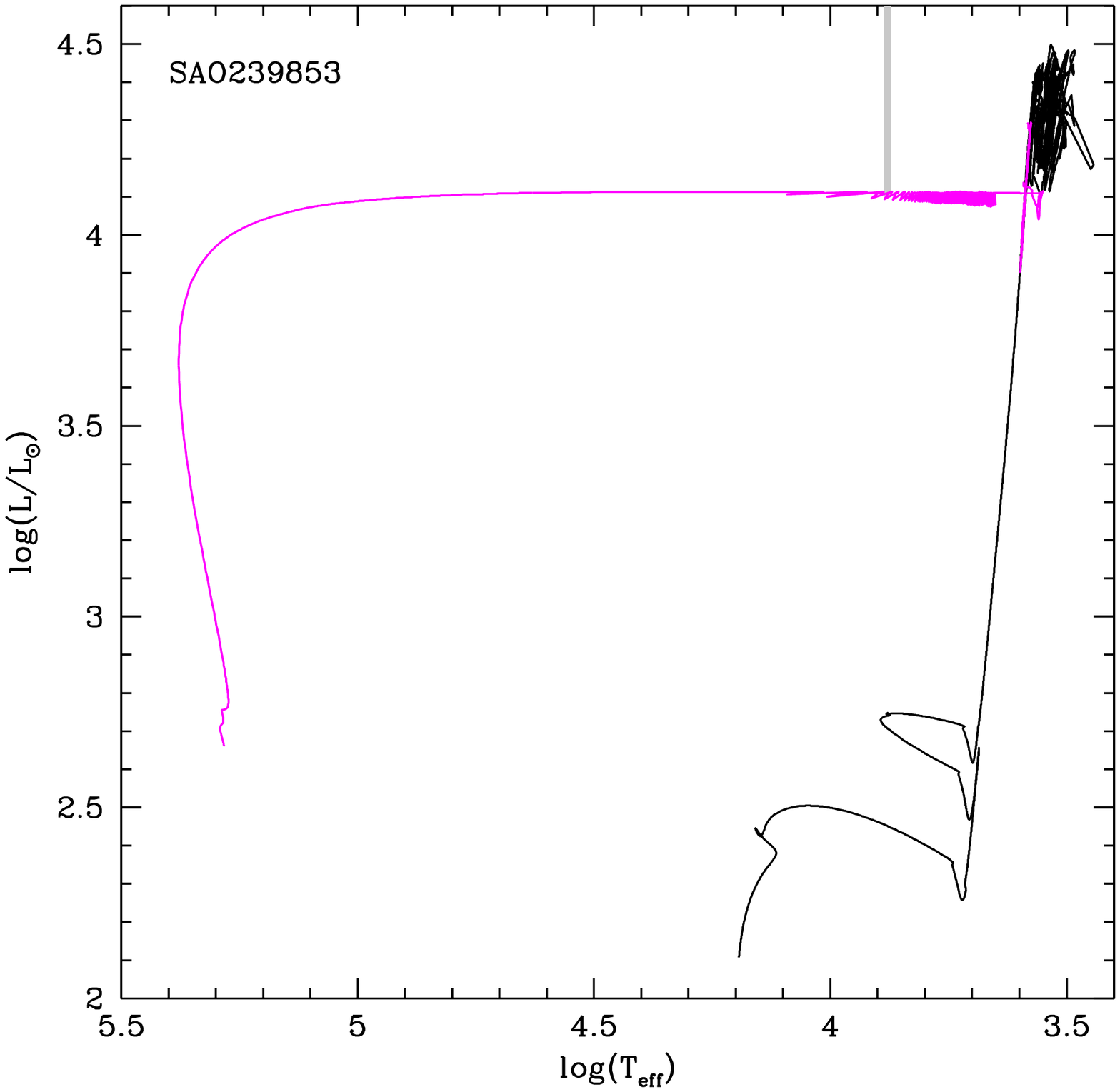}}
\end{minipage}
\vskip-40pt
\caption{Left: AGB variation of the luminosity of stars of metallicity $Z=2\times 10^{-3}$ and initial mass
$2.5~{\rm M}_{\odot}$ (blue line), $3~{\rm M}_{\odot}$ (green), $4~{\rm M}_{\odot}$ (black) and $5~{\rm M}_{\odot}$ (red); the grey-shaded box indicates the luminosity of SAO 239853, with the relative uncertainties.
Middle: Evolution of the surface carbon and nitrogen of the same model stars shown in the left panel, but the
$2.5~{\rm M}_{\odot}$ case; grey shading refers to the chemical composition of SAO 239853. Right: 
evolutionary track of the $3~{\rm M}_{\odot}$ model star, compared with the data of the effective
temperature and luminosity of SAO 239853.}
\label{fsao}
\end{figure*}

\subsubsection{SAO 239853}
SAO 239853 is among the brightest sources in the sample considered, with an uncertain luminosity, given in the
$13000 - 48500~\rm{L}_{\odot}$ range. As shown in the left panel of Fig.~\ref{fsao}, these luminosities
indicate $\rm{M} \geq 3~\rm{M}_{\odot}$ progenitors. Indeed $2.5~\rm{M}_{\odot}$ stars are expected to 
enter the post-AGB phase with luminosities within the limits given above, but they would experience a
large number of TPs and TDU events, which is at odds with the lack of s-process enrichment and the
surface carbon observed, only slightly super-solar. We will therefore consider 
$\rm{M} \geq 3~\rm{M}_{\odot}$ stars in the following, as possible progenitors of SAO 239853.

In the middle panel of Fig.~\ref{fsao} we show the variation of the surface carbon and nitrogen during the
AGB phase of model stars of initial mass $3, 4, 5~\rm{M}_{\odot}$. The effects of the various mechanisms
able to alter the surface chemistry are outline in the figure. The grey zone indicates the observed values.
These results rule out the $4~\rm{M}_{\odot}$ and $5~\rm{M}_{\odot}$ cases. Indeed in these model stars HBB is
very strong since the initial AGB phases, which triggers a fast decrease in the surface carbon, along with
a significant synthesis of nitrogen, which rises up to $[$N$/$Fe$] \sim 1$. During the second part of the
AGB phases the effects of TDU restore the initial carbon, consistently with the observations; however,
the availability of primary carbon causes a further synthesis of nitrogen, which grows to
$[$N$/$Fe$] > 1$, much higher than the observed value. 

We consider the $3~\rm{M}_{\odot}$ possibility, as the $3~\rm{M}_{\odot}$ model star evolves to
surface carbon and nitrogen abundances consistent with those observed for SAO 239853 during the first part of the AGB phase, after the star experienced a couple of TDU events, sufficient raise the surface carbon from the
post-RGB to the observed values. We artificially removed the envelope of the stars from this point on, and
obtained the evolutionary track shown in the right panel, with the uncertainties related to the observed values.

\begin{figure}
\begin{minipage}{0.48\textwidth}
\resizebox{1.\hsize}{!}{\includegraphics{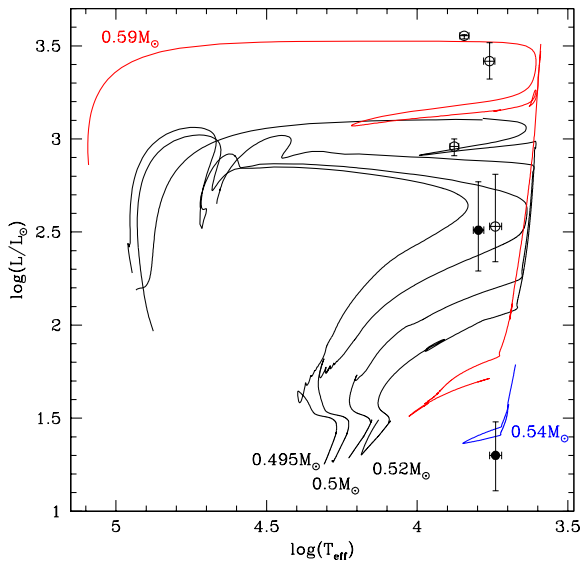}}
\end{minipage}
\vskip-60pt
\caption{Effective temperature and luminosity of the faint sources belonging to the sample presented in paper 1 (from brightest to faintest): HR 7671, IRAS 08187-1905, HD 107369, IRAS 01259+6823, IRAS 05341+0852, IRAS 07430+1115. Black tracks refer to the HB and post-HB evolution of $0.495~{\rm M}_{\odot}-0.52~{\rm M}{\odot}$ stars, that start contraction before reaching the AGB; red line refers to a $0.59~{\rm M}{\odot}$ star, that evolves to the blue after experiencing
two thermal pulses; blue track indicates the core helium burning evolution of a $0.54~{\rm M}{\odot}$ star.
The given masses refer to the beginning of the HB phase.
}
\label{figfaint}
\end{figure}

\subsection{Towards the faint side: HB and post-HB stars}
The sample presented in paper I includes 6 sources whose luminosity is not consistent with the hypothesis that they have undergone a complete AGB evolution and are now contracting to the PN phase: IRAS 05341+0852, IRAS 08187-1905, HD 107369, IRAS 01259+6823 IRAS 07430+1115, HR 7671. Indeed the luminosities are definitively below $3800~\rm{L}_{\odot}$, thus inconsistent with the values reported in tab. 1, even in the very low-mass domain. The position of these stars in the HR diagram is reported in Fig.~\ref{figfaint}.

The existence of low-luminosity stars likely contracting  towards the white dwarf stage was first noticed by \citet{devika16}, who discovered a sample of stars in the Magellanic Clouds, surrounded by dust, with luminosities in the $200-2000~{\rm L}_{\odot}$ range: the interpretation proposed by the authors was that those stars are currently  evolving off the RGB, as a result of binary interaction, which favoured an early loss of the external mantle. A similar explanation can be hardly applied to the present case, as there are no indications that the 4 above sources  belong to binary systems. Indeed we are not considering at all the possibility that these stars are evolving off the RGB as a consequence of the loss of the envelope, even as single stars. Three out of the 4 stars have luminosities below $1000~{\rm L}_{\odot}$, whereas we expect that most of the mass loss during the RGB evolution of single stars, which might potentially lead to an early contraction before the occurrence of the
helium flash, would take place during the approach to the tip of the RGB, thus at luminosities close to $2000~{\rm L}_{\odot}$. For what concerns HR 7671, it is much brighter than the RGB tip, thus it is definitively evolving through a post-RGB phase. 

For 5 of these stars, with the sole exception of IRAS 07430+1115, we propose that they descend from low-mass progenitors, that concluded the horizontal branch (hereafter HB) evolution, and are currently evolving through a post-HB phase, started after the exhaustion of helium in the core. These are the oldest objects in the sample, that formed $6-10$ Gyr ago; unfortunately no robust derivation of the age of the stars can be obtained here, as we will see that our analysis allows an estimate of the mass at the beginning of the HB, but unfortunately the initial mass of the star, the crucial quantity to infer the formation epoch, depends on the mass loss occurred during the RGB, which can change significantly from star to star.

It is clear from Fig.~\ref{figfaint} that the luminosity range of these sources extends over almost 2 orders of magnitude, that reflects the different progenitor mass\footnote{In this case we refer to the mass at the beginning of the quiescent core helium burning phase} and the evolutionary phase during which they are currently observed. 
We will discuss the different cases separately.

\subsubsection{AGB-manqu\'e stars}
In stars of initial mass below $\sim 2~\rm{M}_{\odot}$
the ignition of helium in the central regions takes place in conditions of electron degeneracy, via a thermally unstable process, known as flash. After the occurrence of the flash the electron degeneracy in the core is removed within $\sim 1$ Myr, and the stars resume core helium burning under conditions of thermal stability, then evolve through the AGB, after the exhaustion of central helium.

An exception to this general behaviour is found when the mass of the envelope above the helium core at the
beginning of the HB phase is reduced to a few hundredths of solar mass: under these conditions
the stars barely reach the AGB, rather after the HB evolution, and a relatively short expansion phase, they start contracting, evolving through the so called AGB-manqu\'e phase \citep{greggio90}. 

To test the possibility that some of the stars discussed here are nowadays in the AGB-manqu\'e phase we calculated various evolutionary sequences, started from the HB, of stars of mass in the $0.49-0.6~{\rm M}_{\odot}$ range, extended until the start of the WD cooling. Some of these tracks are shown in Fig.~\ref{figfaint}. 

We find that the AGB-manqu\'e evolution is experienced by $0.49~{\rm M}_{\odot} \leq {\rm M} \leq 0.52~{\rm M}_{\odot}$ stars, which correspond to structures made up of a helium core and an envelope with mass ${\rm M}_{\rm env}$ below $0.04~{\rm M}_{\odot}$. More specifically, for ${\rm M}_{\rm env} < 0.01~{\rm M}_{\odot}$ the evolutionary tracks move vertically then turn to the blue (this case is not shown in Fig.~\ref{figfaint} for clarity sake), whereas for $0.01~{\rm M}_{\odot} < {\rm M}_{\rm env} < 0.04~{\rm M}_{\odot}$ the stars begin the overall contraction after evolving to the red side of the diagram; this behaviour is followed by the model stars
described by the black lines in Fig.~\ref{figfaint}. Note that the total masses given above might change according to different factors that affect the determination of the core mass found at the tip of the RGB, which is core mass assumed for the computation of the core helium burning phase; on the other hand the threshold envelope masses delimiting the different types of post-HB evolutions are substantially independent of metallicity and other physical and chemical ingredients.

In the $0.495M_{\odot}$ and $0.5M_{\odot}$ cases we note the effects of a late helium ignition just during the PN phase, before the WD evolution. The $0.52M_{\odot}$ model star experiences a single thermal pulse, which makes the evolutionary track to describe a wide loop to the blue, before moving again to the red, before the final contraction, preceding the WD evolution.

The effective temperatures and luminosities of IRAS 05341+0852, IRAS 01259+6823 and HD 107369 are fully consistent with the evolutionary tracks of $0.5-0.52~{\rm M}_{\odot}$ stars, thus we suggest an AGB-manqu\'e characterisation for both sources. According to this understanding these stars started the HB phase with envelope masses $\sim 0.02~M_{\odot}$ (IRAS 05341+0852 and IRAS 01259+6823) and $\sim 0.04~M_{\odot}$ (HD 107369), thus descending from hot HB stars, similar to those populating the blue side of the HBs of some Galactic Globular Clusters, such as NGC 2419 \citep{ripepi07} and NGC 2808 \citep{bedin00}.

\subsubsection{An extremely short AGB phase}
HR 7671 and IRAS 08187-1905 are the brightest sources which failed to complete (or to reach) the AGB phase. Their luminosity is however higher than the two sources examined earlier in this section, which is inconsistent with the
hypothesis that it is evolving in the post-HB phase. This can be seen in Fig.~\ref{figfaint}, where we
notice that the evolutionary tracks of  $0.49-0.52~\rm{M}_{\odot}$ stars are too faint to explain HR 7671
and IRAS 08187-1905.

The position of these sources in the HR diagram is nicely reproduced by the evolutionary track of the $0.59~\rm{M}_{\odot}$ model star. We therefore propose that they are the slightly higher mass counterparts of IRAS 05341+0852, IRAS 01259+6823 and HD 107369: the mass loss experienced during and shortly after the HB evolution was not sufficient to remove the whole envelope, thus HR 7671 and IRAS 08187-1905 could reach the AGB phase and started contracting shortly after the beginning of the thermal pulses phase. 

\citet{greggio90} proposed that the evolution of these objects is halted at the first TP, as a consequence
of the sudden increase in the luminosity of the star. We prefer to keep more general in the present context, and propose that the mass of the envelope at the end of the HB phase was so small that the contraction began shortly after the start of this evolutionary phase. Numerical experiments in which the mass of the envelope was artificially removed at the ignition of the first thermal pulse show that within the context of the scenario proposed by \citet{greggio90} the mass of HR 7671 and IRAS 08187-1905 at the beginning of the HB would be $\sim 0.62~{\rm M}_{\odot}$.

Independently of the details of the evolutionary stage at which the residual mass of the envelope is completely removed from the star and contraction begins, the above arguments suggest that HR 7671 and IRAS 08187-1905 descend from HB stars originally composed by the helium core and a $\sim 0.1~{\rm M}_{\odot}$ envelope, which shortly after the beginning of the AGB phase started the overall contraction that led to the current evolutionary stage.

\subsubsection{A core helium burning star?}
Among the low-luminosity stars considered here, IRAS 07430+1115 is the most difficult to interpret, because the luminosity, in the $12-30~{\rm L}_{\odot}$ range, is largely inconsistent with the post-HB hypothesis, claimed for the other 3 sources. Based on the derived effective temperature and luminosity, we suggest that this source is a $\sim 0.55~{\rm M}_{\odot}$ star, currently undergoing core helium burning. Indeed we see in Fig.~\ref{figfaint} that the HB evolution of the $0.54~{\rm M}_{\odot}$ model star is characterised by temperatures and luminosities consistent with those derived for IRAS 07430+1115. For readability we report in Fig.~\ref{figfaint} only the evolutionary track of the $0.54~{\rm M}_{\odot}$ case, but consistency with the parameters of IRAS 07430+1115 is found for all the stars with HB masses in the $0.53-0.55~{\rm M}_{\odot}$ range.

The $0.54~{\rm M}_{\odot}$ star evolves at the effective temperature of IRAS 07430+111 both shortly after the HB evolution begins and towards the end, when the central helium drops below $10\%$. We are more favourable to the
first option, because the spectral energy distribution of the star exhibits a significant infrared excess,
evidence of the presence of dust, likely formed during the RGB ascending: we believe plausible that the dust survived until the beginning of the HB phase, while it would be hard to explain how the dust around the star would not be dispersed after $\sim 10^8$ yr, the time required for the central helium to drop to $10\%$.

\subsubsection{The presence of lithium and carbon enrichment}
The classification proposed for the low-luminosity stars was based on the evolutionary properties and
the evolutionary phase they are evolving through. On the side of the surface chemical composition
we can broadly distinguish two groups: IRAS 05341+0852 and IRAS 07430+1115 show up lithium in the
atmosphere and evidence of carbon and s-process enrichment, whereas HD 107369 and HR 7671 are
characterised by $[$C$/$Fe$]<0$ and no lithium.

The presence of lithium in the atmosphere was shown to be rather frequent in clump stars undergoing helium 
burning, as recently outlined by \citet{kumar20}, based on GALAH and Gaia data. This is ad odds with standard
evolution modelling computations, that find severe lithium depletion during the RGB evolution of low-mass stars,
eventually leading to the complete destruction of the surface lithium. The stars are therefore expected to enter
the HB phase with basically lithium-free envelopes. 

A possible way of escaping this difficulty, recently proposed by \citet{schwab20}, is the occurrence of 
extra-mixing during the helium flash, such that the convective zone formed as a consequence of the activation of the 
helium-burning layer and the external envelope partially overlap, in such a way that the \citet{cameron}
mechanism is activated and lithium synthesis takes place. \citet{schwab20} suggests that this mixing is stimulated by 
internal gravity waves, excited by turbulent convective motions, when the first flash event takes place\footnote{The 
helium flash terminology commonly refers to the whole unstable helium burning phase, which in turn is split into a series 
of flash events, the first being the most energetic.}. 

Understanding the physical nature of this mixing is far beyond the purposes of the present investigation. What is 
interesting for the present study is that a significant fraction of low-mass stars undergoing helium
flash experience a mixing episode, which leads to the synthesis of lithium. The depth of this mixing, i.e.
the extent of the overlap between the internal convective region and the external mantle of the star, is 
still debated, and possibly changes from star to star. If the mixing is sufficiently deep the surface convection 
is mixed with layers exposed to helium nucleosynthesis, thus enriched in carbon: this would favour a
significant increase in the surface carbon, along with the lithium enrichment. The surface lithium and
carbon enrichment of IRAS 05341+0852 and IRAS 07430+1115 suggest that these two sources experienced this
kind of mixing at the ignition of the helium flash.

Further modelling of the helium flash of low-mass stars is required before this understanding can be
confirmed. We leave this problem open.

\section{Conclusions}
We used stellar evolutionary computations, extended until the white dwarf cooling phase, of $0.7-4~{\rm M}_{\odot}$ 
stars with metallicities from $Z=10^{-3}$ to solar, to investigate a sample of post-AGB stars in the Galaxy, for 
which the surface chemistry and the luminosities, based on Gaia parallaxes, have been recently presented. The knowledge of the
luminosity proves crucial for the study of this class of stars, given the tight connection between the
luminosity, the current mass of the stars and the mass of the progenitors. 

Among the various observables, we find that the combined knowledge of the surface carbon and of the luminosity 
proves the best indicator of the past history of the individual sources and of the nature of their
progenitors. 

The comparison with results from stellar evolution modelling shows that $\sim 40\%$ of the
stars in the sample examined descend from $1-3~{\rm M}_{\odot}$ progenitors, that after repeated convective events 
became carbon stars: these objects are s-process enriched and exhibit a surface C$/$O above unity. 

5 sources are the progeny of low-mass stars, that started the AGB
phase with mass below $\sim 1~{\rm M}_{\odot}$, and began the contraction
to the post-AGB before experiencing any carbon enrichment: according to this understanding these objects, with
luminosities in the $4000-5000~{\rm L}_{\odot}$ range, are the oldest stars in the sample.

The three brightest stars, whose surface chemical composition shows up the signature of proton-capture processing, are identified as 
the youngest stars in the sample, descending from $3-4~{\rm M}_{\odot}$ progenitors that experienced both third dredge-up 
and hot bottom burning.

A few sources are characterised by luminosities far below the minimum threshold of $4000~{\rm L}_{\odot}$ 
expected for stars evolving through the AGB. These objects are tentatively identified as the progeny of low-mass 
($\sim 0.5-0.7~{\rm M}_{\odot}$), post core helium burning stars, which after a short expansion phase lost the 
entire envelope and failed to reach the AGB.

The present study confirms that post-AGB stars are a precious tool to reconstruct the evolution of the stars 
through the post-MS phases. The analysis of the chemical composition of post-AGB stars of different mass can potentially shed
new light on still poorly constrained aspects of the AGB evolution, a few examples being the mass transition separating 
low-mass objects evolving as M-type stars from carbon stars, the mass and luminosity threshold required for
the ignition of hot bottom burning, the possibility that stars experiencing HBB eventually get enriched
in carbon owing to late third dredge-up events, deep mixing during the red giant branch pase. The availability 
of Gaia parallaxes is crucial to this research, because it allows the study of the vast sample of Galactic post-AGB, 
so far hampered by the poor knowledge of the distances, that prevented the determination of one essential ingredient 
for this study, namely the luminosity of the individual objects.

Future incoming developments of this study will be a more detailed analysis of the stars considered in the present
work, due to the tighter constrains on the luminosity that will come from DR4, and the application to these key
concepts to a wider post-AGB Galactic sample.

\begin{figure*}
\begin{minipage}{0.48\textwidth}
\end{minipage}
\begin{minipage}{0.48\textwidth}
\end{minipage}
\vskip-80pt
\begin{minipage}{0.48\textwidth}
\end{minipage}
\begin{minipage}{0.48\textwidth}
\end{minipage}
\vskip-60pt
\caption{}
\label{fccd}
\end{figure*}







\bsp	
\end{document}